\documentclass[12pt]{article}
\pdfoutput=1

\newif\ifarxiv
\arxivtrue

\ifarxiv
\def\figpath{.}
\else
\def\figpath{./HLT-diagrams}
\fi

\usepackage[T1]{fontenc}
\usepackage[latin1]{inputenc}
\usepackage{amsmath,amsfonts,amssymb}
\usepackage{setspace}
\usepackage[bookmarks=true,hyperfigures=true,bookmarksnumbered,hidelinks]{hyperref}
\usepackage{graphicx}
\usepackage{sidecap}
\usepackage{tocloft}
\usepackage{tcolorbox}
\usepackage[shortlabels]{enumitem}
\usepackage{cite}
\usepackage{tikz}
\usetikzlibrary{decorations.markings,arrows}

\usepackage[a4paper,text={175mm,257mm},centering]{geometry}

\allowdisplaybreaks[3]
\numberwithin{equation}{section}

\makeatletter
\renewcommand\section{\@startsection {section}{1}{\z@}%
{-3.5ex \@plus -1ex \@minus -.2ex}%
{2.3ex \@plus.2ex}%
{\normalfont\large\bfseries}}
\renewcommand\subsection{\@startsection{subsection}{2}{\z@}%
{-3.25ex\@plus -1ex \@minus -.2ex}%
{1.5ex \@plus .2ex}%
{\normalfont\normalsize\bfseries}}
\makeatother

\expandafter\def\expandafter\bfseries\expandafter{\bfseries\ifmmode\else\boldmath\fi}
\expandafter\def\expandafter\mdseries\expandafter{\mdseries\ifmmode\else\unboldmath\fi}
\expandafter\def\expandafter\normalfont\expandafter{\normalfont\ifmmode\else\unboldmath\fi}

\providecommand{\href}[2]{#2}
\newcommand{\arxivlink}[1]{\href{http://arxiv.org/abs/#1}{[arXiv:#1]}}

\let\oldbib=\thebibliography
\def\thebibliography{\phantomsection\addcontentsline{toc}{section}{\refname}\oldbib}

\let\oldtoc=\tableofcontents
\def\tableofcontents{\phantomsection\addcontentsline{toc}{section}{\contentsname}\oldtoc}

\newcommand{\mathsym}[1]{{}}
\def\id{\protect{{1 \kern-.28em{\rm l}}}}
\def\be{\begin{eqnarray}}
\def\ee{\end{eqnarray}}

\def\tr{{\rm tr}}
\def\ha{\tfrac{1}{2}}
\def\td{\tilde}
\def\ci{\cite}

\def\sm{\sm\ }

\def\a{\alpha}
\def\b{\beta}

\def\g{\gamma}

\def\k{\kappa}

\def\l {\lambda}

\def\O{{\mathcal O}}
\def\m{\mu}

\def\e{\epsilon}
\def\foot{\footnote}
\newcommand{\rf}[1]{(\ref{#1})}

\def\no{\nonumber}

\def\la{\label}
\def\l{\lambda}

\def\p{\phi}
\def\r{\rho}

\def\varpi{{\rm w}}

\def\La{{\Lambda}}

\def\del{\partial}
\def\s{\sigma}
\def\eps{{\epsilon}}
\def\n{\nu}

\def\ed{\end{document}}

\def\iffa{\iffalse}
\def\xp{x_+}
\def\xm{x_-}

\def\ep{\epsilon}

\def\dx{\dot x}

\def\tx{{\tilde x}}

\def\d{\delta}

\def\L{\mathcal{L} }
\def\pcm {{\rm PCM}}
\def\wz{{\rm WZ}}
\def\pcmq {{\rm PCM}$_q$\ }
\def\ZMq {{\rm ZM}$_q$\ }

\def\pp{{\rm p}}
\def\qq{{\rm q}}

\def\rrk{{\rm e}}
\def\ss{{\rm s}}
\def\tt{{\rm t}}
\def\uu{{\rm u}}
\def\sm{$\sigma$-model }
\def\sms{$\sigma$-models }
\def \ov {\over}
\def\Ad{\text{Ad}}
\def\lmp{\Lambda}

\begin{document}

\begin{flushright}\small{Imperial-TP-AT-2018-{05}}
\end{flushright}

\vspace{2.5cm}

\begin{center}

{\Large\bf On the massless tree-level S-matrix \\
\vspace{0.3cm}
in 2d sigma models}

\vspace{1.5cm}

{Ben Hoare$^{a,}$\footnote{bhoare@ethz.ch}, Nat Levine$^{b,}$\footnote{n.levine17@imperial.ac.uk} and
Arkady A. Tseytlin$^{b,}$\footnote{Also at Lebedev Institute and ITMP, Moscow State University. \ tseytlin@imperial.ac.uk}
}

\vspace{0.5cm}

{\em
\vspace{0.15cm}
$^{a}$ETH
Institut f\"ur Theoretische Physik, ETH Z\"urich,\\
\vspace{0.05cm}
Wolfgang-Pauli-Strasse 27, 8093 Z\"urich, Switzerland.
\\
\vspace{0.15cm}
$^{b}$Blackett Laboratory, Imperial College, London SW7 2AZ, U.K.
}
\end{center}

\vspace{0.5cm}

%%%%%%%%%%%%%%%%%%%%%%%%%%%%%%%%%%%%%%
\begin{abstract}
\noindent
Motivated by the search for new integrable string models, we study the
properties of massless tree-level S-matrices for 2d $\s$-models expanded near
the trivial vacuum. We find that, in contrast to the standard massive case,
there is no apparent link between massless S-matrices and integrability: in
well-known integrable models the tree-level massless S-matrix fails to
factorize and exhibits particle production. Such tree-level particle
production is found in several classically integrable models: the principal
chiral model, its classically equivalent ``pseudo-dual'' model, its non-abelian
dual model and also the $SO(N+1)/SO(N)$ coset model. The connection to
integrability may, in principle, be restored if one expands near a non-trivial
vacuum with massive excitations. We discuss IR ambiguities in 2d massless
tree-level amplitudes and their resolution using either a small mass parameter
or the $i\epsilon$-regularization. In general, these ambiguities can lead to
anomalies in the equivalence of the S-matrix under field redefinitions,
and may be linked to the observed particle production in integrable models. We
also comment on the transformation of massless S-matrices under $\s$-model
T-duality, comparing the standard and the ``doubled'' formulations (with
T-duality covariance built into the latter).
\end{abstract}
%%%%%%%%%%%%%%%%%%%%%%%%%%%%%%%%%%%%%%

\newpage
\setcounter{footnote}{0}
\setcounter{section}{0}

\tableofcontents

%%%%%%%%%%%%%%%%%%%%%%%%%%%%%%%%%%%%%%
\section{Introduction}\label{intro}
%%%%%%%%%%%%%%%%%%%%%%%%%%%%%%%%%%%%%%

Examples of integrable 2d models are few and hard to find. In the context of
string theory, one is interested in finding new classically
integrable 2d $\s$-models (see, e.g., \cite{Wulff:2017vhv} and refs. there).
As a direct search for a Lax pair is generally complicated without additional
clues such as symmetries, one may hope that the study of (classical)
S-matrix may provide a useful guide. Indeed, a standard strategy in
construction of {\it massive} integrable 2d theories is to require that the
S-matrix satisfies the conditions of no particle production and factorization %v3
\cite{zam,dor,Parke:1980ki}.\foot{The existence of higher conserved charges implies
(i) the absence of particle production and (ii) equality of sets of initial and final momenta.
Combined with locality and causality this further implies (iii) factorization of $n \to n$ amplitudes into
products of $2 \to 2$ ones.}
Expanded near a trivial flat vacuum, the \sm
$\L=
(G_{mn} + B_{mn} ) \del_+ x^m \del_- x^n = ( \delta_{mn} + h_{mnk} x^k +
c_{mnkl} x^k x^l + \dots) \del_+ x^m \del_- x^n $ describes an interacting theory
of a set of {\it massless} 2d scalar fields. If integrability were equivalent
to the factorization of their S-matrix, one could in principle use it to
determine which couplings $h_{mnk}$, $c_{mnkl}$, etc., i.e. which target space
geometries, correspond to integrable models.

As we will demonstrate below, the connection between classical integrability
and the factorization of the tree-level S-matrix breaks down in the case of
massless 2d scalar scattering: well-known integrable models happen to have
non-zero particle-production amplitudes. Thus the absence of tree-level
massless particle production cannot be used as a criterion in search for
integrable models.

To maintain the link to integrability one should instead consider the expansion
near non-trivial vacua where excitations are massive.\foot{More generally, one
may try to consider the scattering of non-trivial massive solitonic states.} To
give a simple example, consider a massive 2d model $\L= \del_+ x^n \del_- x^n -
V(x), \ V= \ha m^2 x^2 + g x^3 + \dots$ with an integrable potential $V(x)$.
This can be generalized to a \sm by adding two ``light-cone'' directions $u,v=
y\pm t$ as\foot{For examples of such models see, e.g., \cite{ppwave} and
refs. there.} $\L \to \hat \L= \del_+ u \del_- v + \del_+ x^n \del_- x^n -
V(x) \del_+ u \del_- u $. Expanded near the trivial vacuum $u=v=x=0$ this \sm
will have non-zero S-matrix elements for any number of massless %v2
$x$-excitations and a non-zero even number of $u$-excitations, which may not,
in general, factorize. At the same time, if we expand near the ``light-cone''
vacuum $u=\tau, \ v=x=0$ then the $x$-excitations will be massive and their %v2
S-matrix will be factorizable for an integrable potential $V$.
This generalizes to other cases, with a familiar example being the expansion near
the BMN geodesic in $AdS_n \times S^n$ models (see, e.g.,
\cite{klose,Wulff:2017vhv}).\foot{Given a \sm model with target space %v2
$M$ that has at least one isometry $u$, one can expand near classical solutions
of the type $u = \tau$ with the remaining fields constant. This can give
masses to some subset of the excitations; however, their interactions will
typically break Lorentz invariance and hence the resulting massive S-matrix
will not be Lorentz invariant. This is the case for the expansion near the BMN
geodesic in $AdS_n \times S^n$ models. Whether the factorization of such an
S-matrix should be correlated with the integrability of the \sm on
$M$ in general is a priori unclear.}

Returning to the perturbative expansion near a trivial ``massless'' vacuum, let
us recall that for massless 2d theories the standard physical interpretation of
the S-matrix may not apply since space is 1-dimensional and particles moving in
the same direction do not separate asymptotically.\foot{Still, massless
S-matrices were formally discussed for integrable theories in the context of a
finite-density TBA \cite{Zamolodchikov:1992zr}. The S-matrix there retains the
interpretation as the relative phase when one particle is moved past another.}
A related issue is the appearance of IR divergences at the quantum level
\cite{Coleman:1973ci}. Despite this, one is certainly able to formally define
the massless S-matrix at the tree level, e.g., from the classical action
evaluated on a solution with special scattering boundary conditions.
Therefore, one may still ask if the resulting massless S-matrix should somehow
reflect the classical (non-)integrability of the theory. This could be
expected given that the standard definition of classical integrability via the
existence of a Lax pair representation of the equations of motion makes no
distinction between the massless and massive cases.

In some early work, this relation between classical integrability and the
massless S-matrix was indirectly called into question. It was found in
\cite{Nappi:1979ig} that the S-matrix of the Zakharov-Mikhailov (ZM) model
\cite{Zakharov:1973pp,Lund:1976ze} exhibits particle production: there are
non-zero tree-level amplitudes with different numbers of incoming and outgoing
particles. This appears to violate the usual intuition from integrability: the
ZM model is classically integrable (admits a Lax pair) since it is classically
equivalent to the principal chiral model (PCM).

Somewhat confusingly, it was taken for granted in \cite{Nappi:1979ig} that
particle production should be absent in the tree-level massless S-matrix of the
PCM, while this was only known to be the case for the non-perturbative massive
S-matrix \cite{zam}. Consequently, the presence of particle production in the
ZM model was interpreted as implying an inequivalence of the PCM and ZM model
at the level of the classical S-matrix.\foot{The two models are, of course,
quantum-inequivalent having opposite 1-loop $\beta$-functions %v2
\cite{Nappi:1979ig}.}

In fact, the standard argument that integrability implies the
absence of particle production and factorization formally applies only to the massive case %v3
\cite{Parke:1980ki,dor}.\foot{In particular, the proof \cite{Parke:1980ki} that the existence of at least two higher conserved charges
implies factorized scattering uses separation of wave packets which is not possible in the massless case.}
Indeed, it was later pointed out in a little-known
work \cite{Figueirido:1988ct} (which was apparently independent of
\cite{Nappi:1979ig}) that the tree level massless S-matrix does exhibit
particle production in classically integrable \sms such as the $S^N$ \sm and
PCM$_q$ (the PCM with a WZ term with coefficient $q$).\foot{The S-matrix
trivialises \cite{Figueirido:1988ct} in the critical WZW case ($q=\pm1$) when
the left and right modes decouple.} It was thus suggested
\cite{Figueirido:1988ct} that, in contrast to what is well-known for massive
theories \cite{Arefeva:1974bk}, these massless scale-invariant 2d theories do
not exhibit a direct relation between classical integrability and factorized
tree-level scattering (or absence of particle production). In turn, demanding
factorization of the tree-level S-matrix may not, in general, be necessary for
the integrability of classical scale-invariant 2d models with massless
excitations.

That this point remains little known and somewhat controversial is illustrated
by the recent work \cite{Gabai:2018tmm}. There, considering a theory of
Hermitian matrix-valued massless fields with 2-derivative interactions, an
alternative definition of particle production was proposed based on the partial
colour-ordered amplitudes, as opposed to the full amplitudes. Imposing the
constraint of no tree-level particle production (in the sense of this
alternative definition) was claimed to lead one directly to the action of the
integrable $U(n)$ PCM. However, it is unclear how to generalize this procedure
to other integrable \sms that do not have a notion of colour-ordered amplitudes
(see Appendix \ref{comments} below).

Massless 2d S-matrices were also discussed recently in non-renormalizable (non
scale-invariant) Nambu-like models \cite{Dubovsky:2012sh,Cooper:2014noa}. In
this case there are no IR divergences (provided each scalar appears in the
action only through its derivative)\foot{Note that this is particular to a
Nambu action for a string moving in flat space, but is not generally so in the
case of a curved target space.} and thus ambiguities related to IR poles appear
to be absent, not only at the tree level but also at the loop level. Here the
(naively expected) relation between factorization of the massless S-matrix and
integrability does appear to hold (with the Nambu action being integrable
beyond tree level only in special critical dimensions).\foot{The corresponding
2d massless S-matrix was suggested to be a useful tool in trying to understand
the world-sheet theory for a confining QCD string \cite{Dubovsky:2013gi}.}

\bigskip

Our aim is to clarify the properties of tree-level massless scattering
amplitudes in bosonic 2d $\s$-models. Our basic examples will be the principal
chiral model (PCM) and models related to it by dualities -- the classically
dual Zakharov-Mikhailov (ZM) model \cite{Zakharov:1973pp} and the non-abelian
dual (NAD) model \cite{Fridling:1983ha,ftd}. We shall also consider the
generalization PCM$_q$ that includes the WZ term, and the classically dual
ZM$_q$ model, as well as the $\lambda$-model \cite{Sfetsos:2013wia} that
interpolates between the WZW model and the non-abelian dual of the PCM. These
models are classically integrable, admitting a Lax pair, as will be reviewed in
section \ref{models}. We will be interested in their tree-level S-matrices in
the trivial vacuum where the basic excitations (taking values in the Lie
algebra) are massless.

As will be discussed in section \ref{kin}, scattering amplitudes of massless
excitations in such scale-invariant 2d theories may have ``0/0'' ambiguities
due to vertices and internal propagators vanishing simultaneously when the
external momenta are taken on-shell. To resolve these ambiguities requires the
use of a particular IR regularization prescription. The two standard ones that
we shall use are the $i\epsilon$-regularization of the massless propagator and
the massive regularization where all massless fields are given the same small
mass, which is set to zero after momentum conservation is imposed and the
amplitude is taken on-shell.

The simplest non-trivial 4-point amplitudes in the above models will be
computed in section \ref{4-pt}. We shall find that despite the classical
equivalence of the PCM$_q$ and ZM$_q$ models their 4-point amplitudes are not
the same -- they differ by an overall coefficient. The same is true also for
the PCM and its path integral dual -- the NAD model.\foot{That %v4
the tree-level 4-point amplitude in the NAD model is different from the one in
the $SU(2)$ PCM was first found in \ci{Subbotin:1995zz} and this disagreement
was interpreted there as being due to IR ambiguities in computing
massless scattering in 2 dimensions.}

Various 5-point and 6-point amplitudes will be computed in section
\ref{higher}, demonstrating that, despite their classical integrability, the
above models exhibit particle production %v3
and/or absence of factorization of their massless S-matrices.
We shall first show that there are non-vanishing $2\to 4$ massless amplitudes in
the PCM and the $SO(N+1)/SO(N)$ coset \sm (confirming an earlier observation in
\cite{Figueirido:1988ct}). The same conclusion will be reached for $2\to3$
amplitudes: they are non-vanishing not only in the ZM model (as originally
found in \cite{Nappi:1979ig}), but also in the NAD model and the PCM$_q$ (with
$q^2\not=1$).

Another context in which the massless 2d scalar S-matrix has been discussed is
2d scalar-scalar (or T-) duality \cite{rt}. One might a priori expect that
S-matrices of two T-dual \sms should be essentially equivalent (directly
related up to a sign flip depending on the numbers of chiral scalars in the
process). This property becomes manifest \cite{rt} in the ``doubled''
formulation of a \sm \cite{t1} where the left and right chiral scalar modes are
represented \cite{fj} by independent off-shell fields. As we shall discuss in
section \ref{Tduality}, the massless S-matrices computed in the standard and
doubled approaches may, in general, differ in a non-trivial way, as they
correspond to different choices of how to resolve the IR ambiguities. In
particular, the S-matrices of two T-dual \sms found in the standard approach
may not agree, reflecting the fact that T-duality transformation is, in
general, non-local and non-linear in fields. This is also what happens for the
non-abelian dual discussed in section \ref{4-pt}.

In section 5 we will summarize the results and comment on a possible
association between massless particle production in integrable models and IR
ambiguities: the resolution of ambiguities may introduce an anomaly of
integrability.

In Appendix \ref{equiv} we shall point out that the IR ambiguities present in
the 2d massless case may lead to potential anomalies in the standard theorem
about the equivalence of S-matrices in two theories related by a general field
redefinition. The doubled formulation of a bosonic \sm will be reviewed in
Appendix \ref{doubled}. In Appendix \ref{comments} we shall explain how the
discussion of massless PCM scattering in terms of partial colour-ordered
amplitudes in \cite{Gabai:2018tmm} is consistent with the non-vanishing
particle production amplitudes found in section \ref{6point}.

%%%%%%%%%%%%%%%%%%%%%%%%%%%%%%%%%%%%%%
\section[Tree-level massless 4-point amplitudes in the principal chiral model and \texorpdfstring{\\}{} related \texorpdfstring{\sms}{sigma-models}]{Tree-level massless 4-point amplitudes in the principal chiral model and related \texorpdfstring{\sms}{sigma-models}}\label{tree}
%%%%%%%%%%%%%%%%%%%%%%%%%%%%%%%%%%%%%%

%%%%%%%%%%%%%%%%%%%%%%%%%%%%%%%%%%%%%%
\subsection{The PCM and related models}\label{models}
%%%%%%%%%%%%%%%%%%%%%%%%%%%%%%%%%%%%%%

We shall use the following conventions. The 2d metric will be
$\eta_{\m\n}=(-1,1)$, \ $\e^{01}=-\e_{01}=1$ and $\del_\pm\equiv
\pm\del_0+\del_1$. The compact group $G$ generators $t_a$ (anti-Hermitian in
the case of $SU(N)$) that satisfy $\left[ t_a, t_b \right] = {f_{ab}}^c t_c $
will have the Killing norm $\g_{ab} = \tr (t_a t_b) = -\frac{1}{2}
\delta_{ab}$. We shall also use the totally antisymmetric constants
$f_{abc}\equiv {f_{ab}}^d \gamma_{dc}= -\ha {f_{ab}}^c$.

The principal chiral model (PCM) is defined by ($\l$ is a coupling constant)
\begin{equation} \label{2.1}
\mathcal{L}_{\rm PCM} = \frac{1}{\l^2} \, \eta^{\m\n} \tr \left( J_\m J_\n \right)=\frac{1}{\l^2} \g_{ab} \eta^{\m\n} J^a_\m J^b_\n \ , \qquad
\qquad J_\m = g^{-1} \del_\m g \ .
\end{equation}
Setting $g= e^{ \l X}, \ X=X^a t_a, $ we get explicitly
\begin{align}
\mathcal{L}_{\rm PCM} &=
\mathcal{L}_0 + \L_\pcm^{(4)} + \L_\pcm^{(6)} +\O(\l^6) \ , \ \ \ \ \ \ \ \ \ \ \mathcal{L}_0= \g_{ab} \del_\m X^a \del^\m X^b = -\ha \del_\m X^a \del^\m X^a \ , \no \\
\L_\pcm^{(4)}
&= - \tfrac{1}{12} \l^2 {f_{ab}}^e f_{cde} X^a X^c \del_\m X^b \del^\m X^d \ ,\la{2.2} \\
\L_\pcm^{(6)} &= \tfrac{1}{360} \l^4 f_{abl} {f_{cm}}^l {f_{dn}}^m {f_{eg}}^n X^b X^c X^d X^e \del_\m X^a \del^\m X^g \ . \la{2.3}
\end{align}
One may generalize PCM to \pcmq by adding the WZ term with an arbitrary
coefficient $q$ (with the ``critical'' cases $q=\pm 1$ corresponding to the WZW
model)
\be \la{pcmq}
\L_{{\rm PCM}_q} = \L_{\rm PCM} + q \L_{\wz} \ ,
\ee
with the equations of motion
\be\la{2.4}
(\eta^{\m\n} + q\, \e^{\m\n} )\del_\m J_\n = 0 \ .
\ee
The leading term in the expansion of the WZ term to be added to \rf{2.2} is
\be \la{2.5}
\mathcal{L}_\text{WZ} = \mathcal{L}^{(3)}_\text{WZ} + \O(\l^3) \ , \qquad \qquad \mathcal{L}^{(3)}_\text{WZ}
= \tfrac{1}{3} \l\, \e^{\mu\nu} f_{abc}\, X^a \del_\mu X^b \del_\nu X^c \ .
\ee

The PCM is classically equivalent to the Zakharov-Mikhailov model (ZM)
\cite{Zakharov:1973pp,Lund:1976ze}, also considered in
\cite{Nappi:1979ig}.\foot{Aspects of such ``pseudodual'' models were also
discussed in \cite{cz}.} Starting with the first-order form of the PCM
equations
\be \la{2.6}
\del_\m J^\m=0\ , \qquad \qquad F_{\m\n}(J) \equiv \del_\m J_\n - \del_\m J_\n + [J_\m, J_\n] =0 \ , \ee
and solving the first equation by introducing a scalar $\p=\p^a t_a $ (with
values in $\text{Lie}(G)$) as $J^\m = \lambda \e^{\m\n} \del_\n \p$, we then
get from the second equation\foot{Here we assume that ZM model has the same
coupling as the PCM. This, in principle, is not required for the classical
equivalence as the coefficient in front of the ZM Lagrangian can be arbitrary.}
\be \la{2.7}
\square \p^a - \ha \l {f_{bc}}^a \e^{\m\n } \del_\m \phi^b \del_\n \phi^c = 0\ ,
\ee
which follows from the Lagrangian
\be
\la{2.8}
\mathcal{L}_{\rm ZM } = \gamma_{ab} \del^\m \p^a \del_\m \p^b +
\tfrac{1}{3} \l \epsilon^{\m\n} f_{abc}\, \p^a \del_\m \p^b \del_\n \p^c \ .
\ee
Note that the interaction term here is the same as the leading contribution to
the WZ term in \rf{2.5}. Indeed, the ZM model \rf{2.8} may be interpreted as a
\sm with flat target space metric and constant $B$-field strength ($\p^a \equiv
X^a$)
\begin{align}
\la{2.9} &
\L = -\ha (G_{ab} \eta^{\m\n} + B_{ab} \e^{\m\n}) \del_\m X^a \del_\n X^b \ , \ \\
&\la{2.10} G_{ab}=\d_{ab}\ , \qquad B_{ab} = - \tfrac{2}{3} \l f_{abc} X^c\ , \qquad \ H_{abc} = -2 \l f_{abc}\ .
\end{align}
One can construct a similar classically equivalent model by starting with
PCM$_q$ model \rf{pcmq}: solving \rf{2.4} as
\be \la{2.11}
J^\m = \lambda (\e^{\m\n} \del_\n \p - q\, \del^\m \p) \ ,
\ee
one finds from the flatness condition in \rf{2.6} the following generalization
of \rf{2.7}, \rf{2.8}
\begin{align}
&\qquad \square \p^a - \ha \l (1-q^2) {f_{bc}}^a \e^{\m\n } \del_\m \phi^b \del_\n \phi^c = 0\ , \la{2.12}
\\
&\mathcal{L}_{{\rm ZM}_q } = \gamma_{ab} \del^\m \p^a \del_\m \p^b +
\tfrac{1}{3}(1-q^2) \l \epsilon^{\m\n} f_{abc}\, \p^a \del_\m \p^b \del_\n \p^c \ . \la{2.13}
\end{align}
Note that this theory becomes free at the WZW points $q^2=1$.\foot{The %v2
limit $q^2=1$ is somewhat subtle. In general, the equation \rf{2.4} or $ (1-q)
\del_- J_+ + (1+q) \del_+ J_- = 0$ is solved by $ J_+ = \ha (1+q) \del_+ \p +
h(\s^+) , \ J_- = - \ha(1-q) \del_- \p + f(\s^-)$. For $q^2\not= 1$ the
arbitrary functions $h$ and $f$ can be absorbed into a redefinition of $\p$.
However, e.g., for $q=1$ the function $f$ is to be kept (as $J_-$ is non-zero for a
generic solution of the WZW equations). Substituting $J_+ = \del_+ \p , \ J_- =
f(\s^-) $ into the equation $F_{-+}=0$ in \rf{2.6} one then gets $ \del_-
\del_+ \p + [f(\s^-), \del_+ \p] = 0$ which is still equivalent to the free
equation $\del_-\del_+ \p=0$ following from \rf{2.13} with $q=1$ after a $\s^-$
dependent rotation of $\p$.}

%v2
While \pcmq is classically equivalent to \ZMq model, the two models are not, in
general, equivalent at the quantum level: as was shown in \cite{Nappi:1979ig}
for $q=0$, the one-loop $\beta$-functions of PCM and ZM are opposite in
sign.\foot{\la{f9}This is easy to see from the expression for the 1-loop
$\beta$-function \cite{bc} for $G_{ab}$ of the general 2d \sm, i.e.
$\beta_{ab} = R_{ab} -\tfrac{1}{4} H_{acd} H_b^{\ cd}$: in the PCM case we get
only the first term contributing while in the ZM case only the second
contribution is present (cf. \rf{2.9}). For general $q$ we get for the
$\beta$-functions of $\l$: $\beta_{{\rm PCM}_q} = \frac{d\l^{-2}}{dt} = C
(1-q^2)$ and $\beta_{{\rm ZM}_q} = \frac{d\l^{-2}}{dt} = - C (1-q^2)^2$ which
match only at the WZW points $q^2=1$.}

One can define a different dual of the PCM known as the ``non-abelian dual
model'' (NAD) by performing the duality transformation at the level of the path
integral, which should ensure the quantum equivalence of the two models
\cite{Fridling:1983ha,ftd}. Starting with the first-order Lagrangian for the
PCM
\be \la{2.14}
\mathcal{L}=\frac{1}{\l^2} \tr \big[ J^\mu J_\mu + \l \e^{\mu\nu}\, Y F_{\mu\nu}(J) \big] \ ,
\ee
where $Y$ is the Lagrange multiplier field imposing the flatness condition on
the current (cf. \rf{2.6}), and integrating out $J_\m$ gives
\be \la{2.15}
\L_{\rm NAD} = \mathbb{G}^{\m\n}_{ab}(Y)\, \del_\m Y^a \del_\n Y^b \ , \ \ \ \ \qquad
(\mathbb{G}^{-1} )_{\m\n}^{ab} =\eta_{\m\n} \g^{ab} + \l\e_{\m\n} {f^{ab}}_c Y^c \ .
\ee
Expanding in powers of $Y$ we get
\be \la{2.16}
\L_{\rm NAD} = \g_{ab} \partial_\m Y^a \partial^\m Y^b- \l \e^{\m\n} f_{abc} Y^a \del_\m Y^b \partial_\n Y^c
- \l^2 {f_{ac}}^h f_{bdh} Y^c Y^d \del_\m Y^a \del^\m Y^b
+ \O(\l^3)\ .
\ee

%v2
The $\lambda$-model \cite{Sfetsos:2013wia} interpolates between the NAD model
and the WZW model. It is constructed by taking the sum of the PCM and WZW
model Lagrangians for group $G$ with fields $\tilde{g}$ and $g$ respectively and gauging
both models with a common gauge symmetry. Fixing the gauge $\tilde{g} = 1$ and
integrating out the gauge field gives the Lagrangian \begin{equation}
\L = \tfrac{2}{\lambda^2(1-\lmp^2)} \tr\big[\tfrac12 g^{-1} \partial_+ g g^{-1} \partial_- g + \text{WZ}(g) + g^{-1} \partial_+ g \frac{\lmp}{1-\lmp\, \Ad_g} \partial_- g g^{-1} \big] \ ,
\end{equation}
where $\lmp$ is the interpolating parameter of the $\lambda$-model,\foot{We use
$\lmp$ to denote this parameter as we have already used $\lambda$ for the
overall coupling of the 2d $\sigma$-models.} $\text{WZ}(g)$ is the standard WZ
term and $\Ad_g$ denotes the adjoint action of $g$, $\Ad_g X = g X g^{-1}$.
The overall coefficient $\tfrac{2}{\lambda^2(1-\lmp^2)}$ is proportional to the
level $k$ of the $\lambda$-model. Setting $g = \exp(\lambda
(1-\Lambda) Y)$ and expanding in powers of $Y$ we find
\begin{equation}\begin{split} \la{lambda}
\L_{\text{$\lambda$-model}} = \g_{ab} \partial_\m Y^a \partial^\m Y^b
& - \tfrac{1+4\lmp+\lmp^2}{(1+\lmp)} \tfrac{\l}{3} \e^{\m\n} f_{abc} Y^a \del_\m Y^b \partial_\n Y^c
\\ &
- (1+10\lmp+\lmp^2) \tfrac{\l^2}{12} {f_{ac}}^h f_{bdh} Y^c Y^d \del_\m Y^a \del^\m Y^b
+ \O(\l^3)\ .
\end{split}\end{equation}
For $\lmp = 0$ we find the WZW model ($q = -1$ in \rf{pcmq}), while $\lmp = 1$
gives the NAD model \rf{2.16}.\foot{The $\beta$-function of $\lambda$ can be
computed using the results of \cite{Appadu:2015nfa} together with the fact that
the level $k$ does not run. Doing so we find $\beta_{\text{$\lambda$-model}} =
\frac{d\lambda^{-2}}{dt} = \frac{4C \Lambda^3}{(1+\Lambda)^2}$, which, as expected,
vanishes at the WZW point $\Lambda = 0$ and agrees with the PCM result, i.e.
$q=0$ in footnote \ref{f9}, at the NAD point $\Lambda = 1$.}

We note that all of the models discussed above in \rf{2.2}, \rf{2.8},
\rf{2.13}, \rf{2.16}, \rf{lambda} have similar structure with global $G$
symmetry acting on the algebra-valued field as $X\to h X h^{-1} $, $h\in G$.
The leading terms in the perturbative expansion are special cases of the
following Lagrangian
\begin{align}
&\mathcal{L}_{\pp,\qq}(X) = \mathcal{L}_0 + \pp \, \mathcal{L}_\text{PCM}^\text{(4)} + \qq \, 	\mathcal{L}^{(3)}_\text{WZ} + \O(\l^3) \no \\
&\ \ \ \ \ = \g_{ab} \partial X^a \partial X^b - \tfrac{1}{12} \pp\l^2 {f_{ab}}^e f_{cde} X^a X^c \del_\m X^b \del^\m X^d
+ \tfrac{ 1}{3} \qq\l \e^{\mu\nu} f_{abc} X^a \del_\mu X^b \del_\nu X^c +\O(\l^3) \ .\la{2.17}
\end{align}
Here $\mathcal{L}_\text{PCM}^\text{(4)}$ and
$\mathcal{L}_\text{WZ}^{\text{(3)}}$ are as in \rf{2.2} and \rf{2.5} and the
coefficients p and q are given in Table 1. Note that in each case we are
studying the perturbation theory expanded around the trivial vacuum $X^a=0$.
\begin{table}
\begin{center}
\begin{tabular}{lcccccccc}
& PCM & PCM$_q$ & WZW & ZM & ZM$_q$ & NAD & $\lambda$-model
\\
p & 1 &1 &1 &0 &0 &12 & {{ $1+10\lmp+\lmp^2$}}
\\
q & 0 & $q$ & $\pm 1$ &1 & $ 1- q^2$ &$-3$ & $- \tfrac{1+4\lmp+\lmp^2}{1+\lmp}$
\end{tabular}
\end{center}
\caption{\small Coefficients p and q in the Lagrangian \rf{2.17} for various models.}
\la{tab1}
\end{table}

All the classically equivalent models discussed above are classically
integrable admitting a flat Lax connection. The Lax connection of the \pcmq
model may be expressed in terms of the components of the current $J_\m$ as ($z$
is the spectral parameter)
\be \la{lax1}
L_+ = \ha( 1 - q + z \sqrt{1 -q^2} ) J_+ \ , \ \ \ \ \ L_- = \ha( 1 + q + z^{-1} \sqrt{1 -q^2} ) J_- \ .
\ee
The equations following from the flatness of the Lax connection are the same as
in \rf{2.4}, \rf{2.6}, i.e.
\be \la{laxeq}
\del_- J_+ - \del_+ J_- + [J_-,J_+] = 0 \ , \qquad
(1-q) \del_- J_+ + (1+q) \del_+ J_- = 0 \ .
\ee
To recover the equation of motion of the \pcmq model we solve the first
equation of \eqref{laxeq} by setting $J_\mu = g^{-1} \del_\mu g$, and
substitute into the second equation.

The first-order form of the equations of motion for the \ZMq model are also
given by \eqref{laxeq}. Therefore, the Lax connection takes the same form
\eqref{lax1}. To recover the equation of motion \rf{2.12} of the \ZMq model we
solve the second equation of \eqref{laxeq} by setting $J_\m \equiv \lambda
(\e_{\m\n} \del^\n \phi -q \del_\m \phi)$, and substitute into the first
equation.

The same Lax connection applies also to the NAD model (setting $q=0$ in
\rf{lax1}) and the $\lambda$-model \cite{Sfetsos:2013wia}. For the NAD model we
now solve a combination of the two equations in \eqref{laxeq} by setting \be
\la{nadsol} J^\m = - \lambda \e^{\m\n}(\del_\n Y + [J_\n,Y]) \ . \ee
Substituting this into the first equation of \eqref{laxeq} gives the equation
of motion of the NAD model.\foot{To see that \eqref{nadsol} solves a
combination of the two equations \eqref{laxeq}, we rewrite it as $\partial_\pm
Y = \pm \lambda^{-1} J_\pm + [Y,J_\pm]$ and substitute into the equation
$\del_- (\del_+ Y) - \del_+ ( \del_- Y) = 0$. After a short amount of algebra
one finds that this is equivalent to $\lambda^{-1} (\del_- J_+ + \del_+ J_-) +
\big[Y, \del_- J_+ - \del_+ J_- + [J_-,J_+]\big] = 0$, which is indeed a
combination of the two equations \eqref{laxeq} for $q=0$.}

%%%%%%%%%%%%%%%%%%%%%%%%%%%%%%%%%%%%%%
\subsection{Comments on massless 2d kinematics}\label{kin}
%%%%%%%%%%%%%%%%%%%%%%%%%%%%%%%%%%%%%%

We shall consider scattering of massless scalar particles in 2 dimensions. In
2d the mass-shell equation $k^2 \equiv - k_0^2 + k_1^2 = 0$ factorizes
as\foot{We shall denote 2d momenta by $k,l, r,v, \dots$. In our conventions
$k\cdot l =\ha (k_+ l_- + k_ - l_+)$.}
\be \la{2.18}
k_+ k_-=0 \ , \qquad \qquad k_\pm \equiv \pm k_0 + k_1 \ .
\ee
Thus the mass-shell consists of two linear solutions $k_+=0$ and $k_-=0$ (``left-moving'' and ``right-moving''), which join at the
special point $k_\m=0$.

\begin{figure}[t!]
\centering
\raisebox{-0.5\height}{\includegraphics[scale=0.7]{\figpath/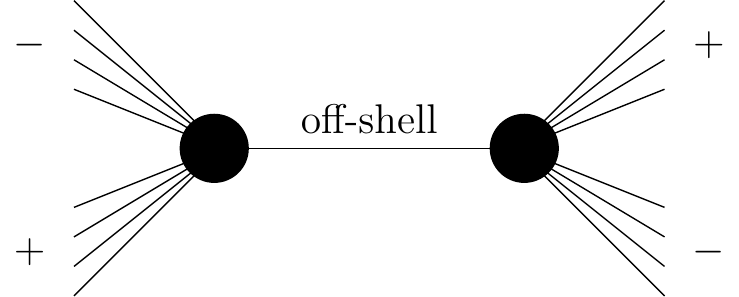}} \hspace{0.5cm} \raisebox{-0.2\height}{\includegraphics[scale=0.7]{\figpath/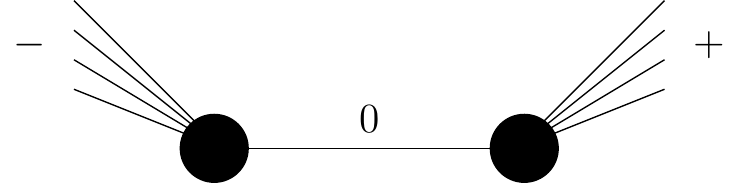}} \hspace{0.5cm} \raisebox{-0.5\height}{\includegraphics[scale=0.7]{\figpath/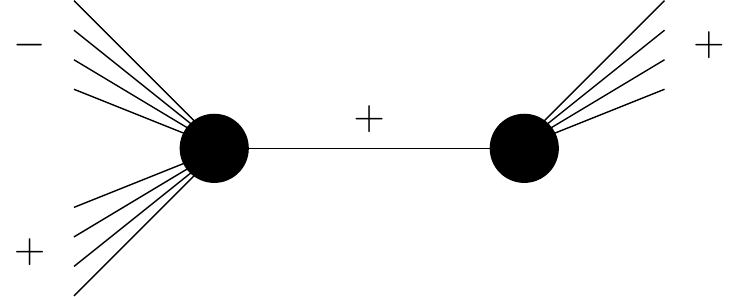}}
\caption[] {\small Different types of divergences in tree diagrams.
The blobs denote some sub-diagrams.
The left case is a generic configuration for which there is no IR divergence.
The middle case depicts a Type 1 divergence with the internal momentum going to zero when external momenta are taken on-shell.
The right case depicts a Type 2 divergence (the $+$ and $-$ labels may be swapped). \label{figambig}}
\end{figure}

The conservation of momentum applies separately to the left- and right-moving excitations (all momenta $k^{(i)}, l^{(j)}$ are incoming here)
\be \la{2.19}
\sum_i k^{(i)}_\m = 0 \ , \qquad \sum_j l^{(j)}_\m =0 \ , \qquad \ \ \ \ \ \ k_+^{(i)} = 0 \ , \ \ l_-^{(j)} = 0 \ .
\ee
The splitting into left- and right-moving excitations with linear mass-shell conditions
leads to two types of divergences when internal propagators blow up --
``Type 1'' and ``Type 2'':
\be \la{2.20}
\text{Type 1: } \ \ \rrk_\m =0 \ ; \qquad \qquad \text{Type 2: } \ \ \rrk^2=0 \ , \ \rrk_\m \neq 0 \ ; \qquad \quad (\rrk = \text{internal momentum})\ \
\ee
Type 1 occurs when, on each side of the propagator, the external momenta are of %v3
same chirality so that due to momentum conservation the components of the internal momentum should both be zero: $\rrk_+=\rrk_-=0$.
Type 2 occurs when the external momenta on just one side of the
propagator are of the same chirality; then $\rrk_+=0$ or $\rrk_-=0$ but, in general, one of them is non-vanishing (see Fig.\ref{figambig}).

In the (classically) scale invariant \sms that will be our focus, the
amplitudes have the potential to remain finite despite these divergences. This
is because each interaction term carries two derivatives, so that every
infinite propagator is compensated by a vanishing vertex factor. Even if all
divergences are compensated by vanishing vertex factors, one may encounter
``0/0'' ambiguities of the form $\frac{V}{\rrk^2}$, where both $V$ and $\rrk^2
$ go to zero as the external legs go on-shell (we shall see examples of this
below).

One possible way to resolve such ambiguities is the standard ``{\it $i
\epsilon$-regularization}'', i.e. the replacement $\frac{V}{ \rrk^2} \to \frac{
V}{\rrk^2-i \epsilon}$ where $\epsilon $ is set to zero only after the external
momenta are taken on-shell (massless). Then the vanishing of $V$ implies that
such ``0/0'' ambiguous contributions should be simply set to zero. This was the
primary approach taken in \cite{Figueirido:1988ct} (and apparently also in
\cite{Nappi:1979ig}).

We will also consider another prescription: {\it ``massive regularization''},
where we introduce a mass term for all the fields in the action with the same
mass parameter $m^2\to 0$. With $m^2 = - i \epsilon$ this is different from
the $i \epsilon$-regularization in that not only the propagators but also the
mass-shell conditions are modified. Explicitly, the massless external momenta
are replaced by massive ones according to the following rules
\be\la{225}
k_+^{(i)} =0\ \ \to\ \ k_+^{(i)} = -\tfrac{m^2}{k_-^{(i)}} \ , \qquad \qquad \qquad l_-^{(j)} =0\ \ \to\ \ l_-^{(j)} = -\tfrac{m^2}{l_+^{(j)}}
\ . \ee
The conservation of momentum in \rf{2.19} then becomes
\be \la{newCOM}
\sum_i k^{(i)}_- = m^2 \sum_j \tfrac{1}{l^{(j)}_+} \ , \qquad \qquad \sum_j l^{(j)}_+ = m^2 \sum_i \tfrac{1}{k^{(i)}_-} \ .
\ee
In order for \rf{newCOM} to be satisfied, the non-vanishing components
$k_-^{(i)}$ and $l_+^{(j)}$ must also be deformed from their $m=0$
values.\foot{One might be concerned that the choice of how to deform these
components leads to an ambiguity. However, it turns out that, as long as one
solves \rf{newCOM} for one $k^{(i)}_-$ and one $l^{(j)}_+$ (in order to obtain
a solution regular as $m\to 0$), and only one mass parameter is used (to avoid
order-of-limits issues), there is no ambiguity. Moreover, the regularity of
this solution and of the amplitude in the $m\to 0$ limit will guarantee that
the result is not dependent on the choice of which variables to eliminate.}

One can immediately see that Type 1 ambiguities vanish in both the massive
regularization and the $i \epsilon$-regularization. In this case the ambiguous
contribution is of the form $\frac{V_1({\rrk}) V_2({\rrk})}{{\rrk}^2}$, with
both $V_1$ and $V_2$ vanishing as the internal momentum goes to zero, i.e.
${\rrk}_\mu \to 0$. In the massive regularization this becomes
$\frac{V_1({\rrk}) V_2({\rrk})}{{\rrk}^2+m^2}$. According to \rf{225},
\rf{newCOM}, all of the would-be vanishing quantities ${\rrk}_\m$, $V_1$ and
$V_2$ will now be of order $m^2$. Hence ${\rrk}^2$ is order $m^4$ and the
ambiguous contribution is vanishing in the $m \to 0$ limit as
\be
\frac{V_1({\rrk}) V_2({\rrk})} {{\rrk}^2+m^2} =\frac{ \O(m^2) \O(m^2) }{\O(m^4) + m^2 } = \frac{ \O(m^4) }{ \O(m^2)} \to 0 \ ,
\ee
in agreement with the $i \e$-regularization.

Let us note that tree-level amplitudes with all particles of the same chirality
vanish in both the massive and $i\ep$-regularizations. Indeed, every vertex
vanishes on-shell so will be of order $m^2$. With only one chirality there are
no Type 1 ambiguities but every internal line is on-shell, producing Type 2
ambiguities. In the $i\e$-regularization these vanish and thus the whole
amplitude vanishes. In the massive regularization they blow up as $m^{-2}$ and
so, in the massless limit, each diagram with ${\rm V}$ vertices and ${\rm L}$
internal lines goes as $A = \O(m^{2{\rm V}}) \times \O( m^{-2{\rm L}})$. Any
tree-level graph has ${\rm V} = {\rm L}+1$ so we get $A= \O(m^2) \to 0$.

%%%%%%%%%%%%%%%%%%%%%%%%%%%%%%%%%%%%%%
\subsection{4-point scattering amplitudes}\label{4-pt}
%%%%%%%%%%%%%%%%%%%%%%%%%%%%%%%%%%%%%%

Our aim will be to compute the simplest tree-level scattering amplitudes for the Lagrangian \rf{2.17} and
thus compare the classical S-matrices for the models listed in Table 1.
We shall be scattering the massless scalars $X^a$ in left-moving and right-moving on-shell states as discussed in section \ref{kin}.

\begin{figure}[t!]
\centering
\raisebox{-0.5\height}{\includegraphics[scale=0.9]{\figpath/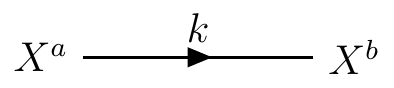}} \hspace{2cm}
\raisebox{-0.5\height}{\includegraphics[scale=0.9]{\figpath/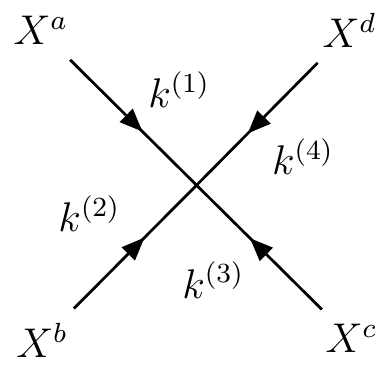}} \hspace{2cm}
\raisebox{-0.5\height}{\includegraphics[scale=0.9]{\figpath/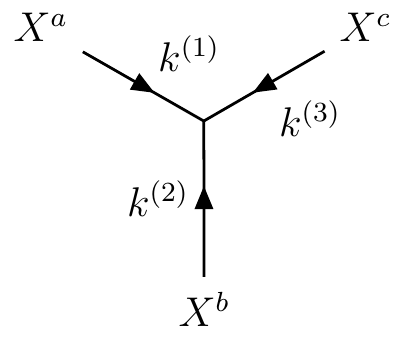}}
\caption[] {\small Feynman rules for the theories listed in Table \ref{tab1}.
\label{fig1}}
\end{figure}

\begin{figure}[t!]
\centering
\raisebox{-0.5\height}{\includegraphics[scale=0.9]{\figpath/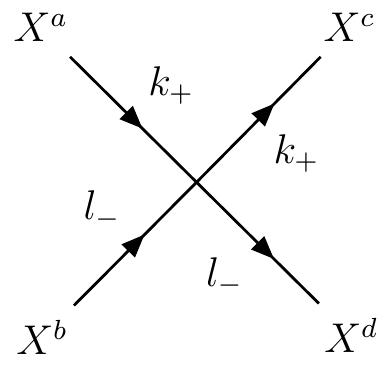}} \hspace{2cm}
\raisebox{-0.5\height}{\includegraphics[scale=0.9]{\figpath/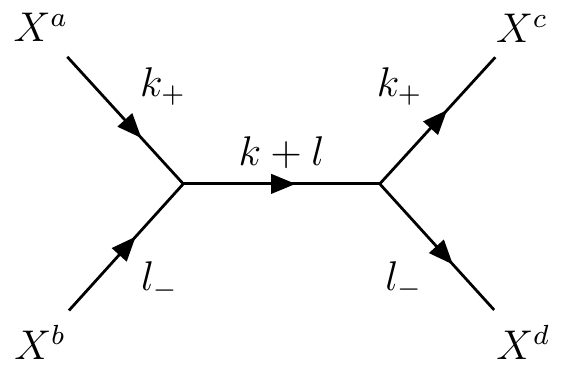}} \hspace{2cm}
\\
\raisebox{-0.5\height}{\includegraphics[scale=0.9]{\figpath/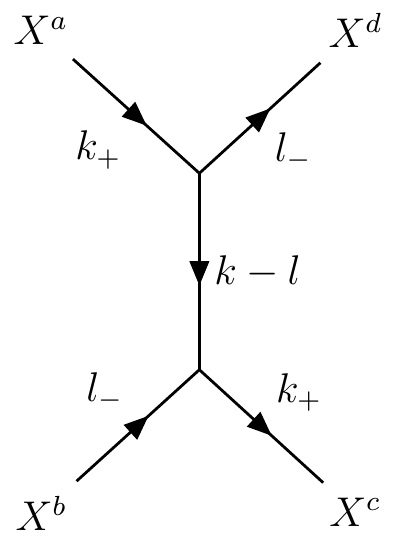}} \hspace{2cm}
\raisebox{-0.5\height}{\includegraphics[scale=0.9]{\figpath/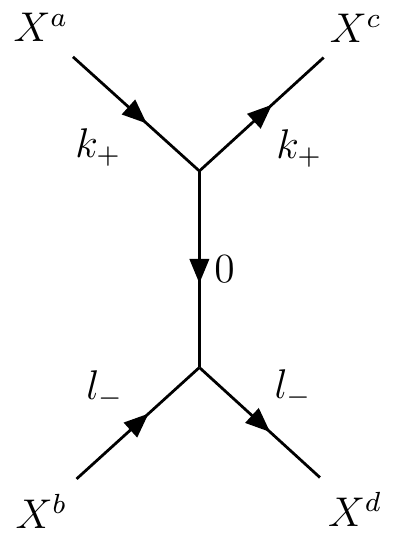}}
\caption[] {\small Contributions to 4-point amplitude. Top left: Contact diagram $A_{\rm cont}$. Top right: Exchange diagram $A^{(\ss)}_{\rm exch}$. Bottom left: Exchange diagram $A^{(\uu)}_{\rm exch}$. Bottom right: Ambiguous exchange diagram $A^{(\tt)}_{\rm exch}$.
\label{fig2}}
\end{figure}

The Feynman rules corresponding to \rf{2.17} are (see Fig.\ref{fig1})\foot{Note that for canonical choice
$\gamma_{ab} = - \ha \delta_{ab}$ the propagator has the standard form $ - \frac{i}{k^2} $.}
\begin{align}
\la{2.21} P^{ab} & = \tfrac{i}{2} \frac{\g^{ab}}{k^2}
\ , \\
V_{abcd} & = \tfrac{i}{6} \pp\, \l^2 \big[{f_{ab}}^m f_{cdm} (k^{(1)}-k^{(2)})\cdot (k^{(3)}-k^{(4)}) + (\text{2 perms}) \big] \ ,
\qquad && \sum_{i=1}^4k^{(i)} = 0 \ \la{2.22} \ , \\
\la{2.23}
V_{abc} & = -2 i \qq\, \l f_{abc} \e^{\m\n} k^{(1)}_\m k^{(2)}_\n \ , && \sum_{i=1}^3k^{(i)} =0 \ .
\end{align}
The 3-point on-shell scattering amplitudes vanish due to massless 2d kinematics
while the non-vanishing 4-point scattering amplitude $+-\to +-$ receives contributions
from the contact PCM 4-vertex in \rf{2.17} and the three exchange diagrams
with the 3-vertices from the WZ-type term in \rf{2.17} (see Fig.\ref{fig2})
\be \la{2.24}
S[X^a (k_+) X^b (l_-) \rightarrow X^c (k_+) X^d (l_-)]
= A_{\rm cont} + A^{(\ss)}_{\rm exch}+ A^{(\tt)}_{\rm exch}+ A^{(\uu)}_{\rm exch} \ .
\ee
Explicitly (suppressing $a,b,c,d$ indices on the l.h.s.) we find
\begin{align}
&\la{2.25} A_{\rm cont}=\tfrac{1}{2^4}
\tfrac{1}{3} i\pp \l^2 \left( f^{abm} {f^{cd}}_m + 2 f^{acm} {f^{bd}}_m + f^{adm} {f^{bc}}_m \right) (k \cdot l)\ , \\
&\la{2.26} A^{(\ss)}_{\rm exch}=
\tfrac{1}{2^4}
\tfrac{i}{2} (2 i \qq \l)^2 ({f^{ab}}_m \e_{\m\n} k^\m l^\n )\frac{\g^{mn}}{(k+l)^2} ( {f^{cd}}_n \e_{\r\s} k^\r l^\s )
= - \tfrac{i}{16} \qq^2 \l^2 f^{abm} {f^{cd}}_m (k \cdot l)\ , \\
&\la{2.27} A^{(\uu)}_{\rm exch}= \tfrac{1}{2^4}
\tfrac{i}{2} (2 i \qq \l)^2 ({f^{ad}}_m \e_{\m\n}k^\m l^\n )\frac{\g^{mn}}{(k-l)^2} ( {f^{bc}}_n \e_{\r\s} l^\r k^\s )
= - \tfrac{i}{16} \qq^2 \l^2 f^{adm} {f^{bc}}_m (k \cdot l)
\ .\end{align}
The $\tt$-channel exchange is an example of a Type 1 divergence in \rf{2.20},
with the internal momentum vanishing when the external legs go on-shell. Indeed, if we
formally replace this amplitude by the corresponding
off-shell (amputated) Green's function $X^a (k)X^b (l) \rightarrow X^c (k-\rrk) X^d (l+\rrk)$, where $\rrk$ is the internal momentum to be set to zero
, we find
\begin{align}
\la{2.28}
A^{(\tt)}_{\rm exch}= &\tfrac{1}{2^4} \tfrac{i}{2} (2 i \qq \l)^2 ({f^{ac}}_m \e_{\m\n} k^\m (-k +\rrk)_\n)\frac{\g^{mn}}{\rrk^2} ( {f^{bd}}_n \e_{\r\s}l^\r (-l-\rrk)^\s )\no\\
= &\tfrac{i}{8} \qq^2 \l^2 f^{acm} {f^{bd}}_m \frac{(\e_{\m\n} k^\m \rrk^\n )( \e_{\r\s} l^\r \rrk^\s)}{\rrk^2} \ , \ \ \ \ \ \ \ \ \rrk_\mu \to 0
\ .\end{align}
As expected, in the on-shell limit $\rrk_\m \to 0$, there are vanishing factors of
equal order in the numerator and denominator of the fraction.

Adding together the four contributions \rf{2.25}, \rf{2.26}, \rf{2.27} and \rf{2.28} and
using the Jacobi identity, one obtains
\begin{align} \la{2.29}
&S[X^a (k_+) X^b (l_-) \rightarrow X^c (k_+) X^d (l_-)] \no \\
&\qquad \qquad= \tfrac{i}{32} \l^2 f^{acm} {f^{bd}}_m \Big[ \left( \pp - \qq^2 \right) \ k_+ l_-
+ 4 \qq^2 \frac{(\e_{\m\n} k^\m \rrk^\n )( \e_{\r\s} l^\r \rrk^\s)}{\rrk^2} \Big] \ , \ \ \ \ \ \ \ \ \rrk_\mu \to 0 \ .
\end{align}
Since the t-channel exchange ambiguity in \rf{2.28}, \rf{2.29} is of Type 1, both the $i\epsilon$-regularization and the massive regularization
resolve it in the same way, giving a vanishing contribution. Hence in both cases the result is
\be \la{2.30}
S[X^a (k_+) X^b (l_-) \rightarrow X^c (k_+) X^d (l_-)] =
\tfrac{i}{32} \l^2 \k \,
f^{acm} {f^{bd}}_m\ k_+ l_- \ , \ \ \ \ \ \ \ \ \ \k\equiv \pp - \qq^2 \ .
\ee
Thus the leading 4-point scattering amplitude for all the theories in Table 1 has this
universal form with the explicit values
of the overall coefficient $\k=\pp-\qq^2$ given in Table 2.

\begin{table}
\begin{center}
\begin{tabular}{lcccccccc}
& PCM & PCM$_q$ & WZW & ZM & ZM$_q$ & NAD & $\lambda$-model \\
$\k $
& 1 &$1-q^2$ &0 &$-1$ &$-(1-q^2)^2$ & 3 & $ \tfrac{4 \lmp (1+\lmp+\lmp^2)}{(1+\lmp)^{2}}$
\end{tabular}
\end{center}
\caption{\small Overall coefficient $\k=\pp - \qq^2$ in the 4-point amplitude in \rf{2.30}.}
\la{tab2}
\end{table}

We find that the $+-\to +-$ tree-level amplitude vanishes in the critical WZW
model.\foot{This was also found earlier in \cite{Figueirido:1988ct} using the
$i\epsilon$-regularization.} This could be expected given the decoupling of the
left-moving and right-moving modes in the classical equations. The same is, of
course, true also for the classically equivalent ZM$_1$ model, which is a free
theory (cf. \rf{2.13}).

We also conclude that the 4-point amplitudes of the classically equivalent PCM
and ZM models are not actually the same -- they differ by an overall sign.
This difference becomes even more substantial for their $q$-generalizations:
the amplitudes of the classically equivalent PCM$_q$ and ZM$_q$ models are
related by $(1- q^2) \to - (1-q^2)^2$.\foot{This is, in fact, the same relation
as in of their 1-loop $\b$-functions (see footnote \ref{f9}): the contact and
exchange contributions in the amplitude have direct counterparts in the Ricci
tensor and the square of the 3-form that enter with the opposite signs in the
$\b$-function.} Moreover, the PCM and NAD models that are related by a path
integral duality transformation and have the same one-loop $\b$-functions
\cite{Fridling:1983ha} also happen to have different tree-level S-matrices.

In fact, the classically equivalent models like PCM, ZM and NAD, whose
classical solutions are in one-to-one correspondence (implying, in particular,
relations between integrable structures or Lax pairs), need not have equivalent
massless S-matrices. One reason is that, while the tree-level S-matrix is
generated by the classical action evaluated on the classical solution with
asymptotic boundary conditions, the classical actions of these models are not
the same. Also, the relation between the elementary scattering fields is
non-local: according to \rf{2.11}, \rf{2.14}, \rf{nadsol}, if $J_\m=e^{-\l X}
\del_\mu e^{\l X}$ then $J^\m =\l \epsilon^{\m\n} \del_\n \phi$ for PCM vs. ZM
and $ J^\m =- \l \epsilon^{\m\n}( \del_\n Y + [J_\n, Y]) $ for PCM vs. NAD.
Moreover, these models have different discrete symmetries: the PCM is
parity-invariant, while the ZM and NAD models contain parity-odd interactions.
As a result, the S-matrices of the latter theories may contain non-vanishing
amplitudes with odd numbers of legs that are automatically absent in the case
of the PCM.

Still, the relation between classical solutions may be suggesting that there
exists some map between the corresponding S-matrix elements. This is supported
by an argument about the duality relation of PCM and NAD in \cite{ftd}.
Introducing a source for the current $J_\m$ in \rf{2.14} and integrating out
$J_\m$ one gets the expression for the generating functional for correlators of
currents in the PCM in terms of the NAD theory path integral. This amounts to
an (off-shell) relation between the correlators of currents in one theory and
the correlators of their counterparts in the dual theory. This should then
also translate into relations between certain on-shell amplitudes.

%v3
%%%%%%%%%%%%%%%%%%%%%%%%%%%%%%%%%%%%%%
\section{Higher-point amplitudes: particle production and non-factorization}\label{higher}
%%%%%%%%%%%%%%%%%%%%%%%%%%%%%%%%%%%%%%

Let us now turn to higher-point scattering amplitudes in the models discussed
in section \ref{models}. Despite the PCM and the classically equivalent ZM and
NAD models being integrable, the corresponding massless S-matrices fail to
factorize and contain non-zero particle production amplitudes.

Thus the standard lore about factorization of the S-matrix of integrable models
does not directly apply to the massless scattering case. This was already
noticed in \cite{Nappi:1979ig} (for the ZM model) and in
\cite{Figueirido:1988ct} (for the $SO(N+1)/SO(N)$ coset model and PCM$_q$ with
$q\neq \pm 1$). Here we shall explicitly confirm this and also find similar
results for the NAD model.

The higher-point amplitudes feature both types of ``0/0'' IR ambiguities
described in section \ref{kin}. In particular, the presence of Type 2
ambiguities will lead, in general, to different results in the
$i\e$-regularization and massive regularization. Below we will mostly use the
massive regularization, as this prescription appears to be better defined %v3
(see Appendix \ref{equiv}).

%%%%%%%%%%%%%%%%%%%%%%%%%%%%%%%%%%%%%%
\subsection{\texorpdfstring{2 $\to$ 4}{2 to 4} amplitudes in \texorpdfstring{$SU(2)$}{SU(2)} PCM and \texorpdfstring{$SO(N+1)/SO(N)$}{SO(N+1)/SO(N)} coset model}\label{6point}
%%%%%%%%%%%%%%%%%%%%%%%%%%%%%%%%%%%%%%

Let us specialize to the PCM for $G = SU(2)$, where $\gamma_{ab} = - \ha
\delta_{ab}, \ f_{abc} = -\ha \e_{abc} \ (a,b=1,2,3)$. Using the massive
regularization we will compute particular
amplitudes $+- \to ---+$ and $+- \to --++$. %v3
The amplitudes with an odd number of left or right particles correspond to particle production.
The amplitudes with an even number, such as $+- \to ---+$ (related by crossing to $+-- \to +--$),
are formally not particle production amplitudes, but in an integrable theory are expected
to be non-zero only when the sets of incoming and outgoing momenta are same and the amplitude factorizes
into product of 2-particle amplitudes. As we shall see below, this will not be so in the present massless case
due to IR ambiguities.

\begin{figure}[t!]
\centering
\raisebox{-0.5\height}{\includegraphics[scale=1.05]{\figpath/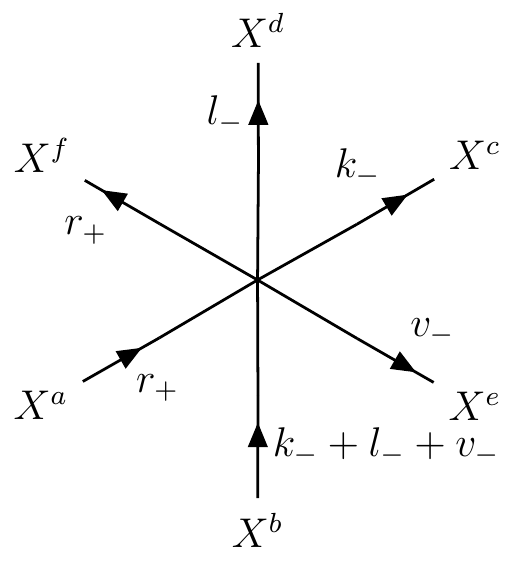}} \\
\raisebox{-0.5\height}{\includegraphics[scale=1.05]{\figpath/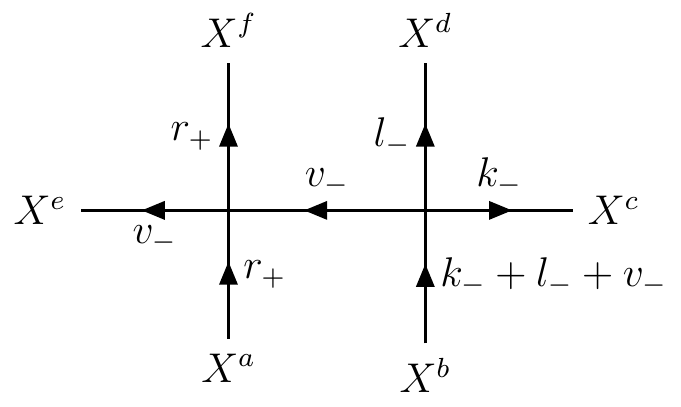}} \hspace{1.8cm} \raisebox{-0.5\height}{\includegraphics[scale=1.05]{\figpath/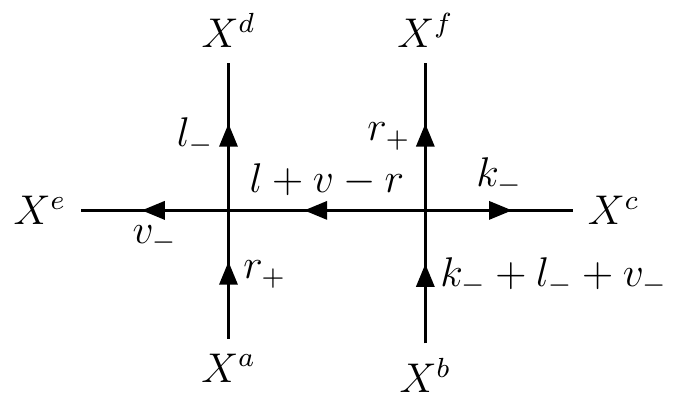}}
\caption{\small Contributions to the 2 $\to$ 4 amplitude \rf{3.10}. Top: Contact diagram $A_{\rm cont}$.
Bottom left: S exchange diagram $A_{\rm exch}^{\rm (S), v}$ contributing to $A_{\rm exch}^{\rm (S)}$. Bottom right: D exchange diagram contributing to $A_{\rm exch}^{\rm (D)}$.
The set of exchange diagrams contains also
the cyclic permutations of the ``$-$'' legs. \label{6fig}}
\end{figure}

In order to illustrate the details of the massive regularization method, we
will explain the computation of the $+- \to ---+$ amplitude in some detail.
According to the Feynman rules \rf{2.2}, \rf{2.3} there is a contact term from
the 6-vertex, as well as the exchange diagrams with two 4-vertices. We may
split up the exchange diagrams into two classes, according to whether the two
``$+$'' legs are incident to the same vertex (S) or to different vertices (D)
(see Fig.\ref{6fig}):
\be\la{3.1}
S[X^a (r_+) X^b (k_- +l_- +v_-) \to X^c (k_-) X^d (l_-) X^e (v_-) X^f (r_+)] = A_{\rm cont} +A_{\rm exch}^{\rm (D)} + A_{\rm exch}^{\rm (S)}
\ . \ee
The contact term is
\begin{align} \la{cont}
A_{\rm cont} &= \tfrac{i}{72} \l^4 r_+ \Big[(k_- - l_-) \d_{ad} \d_{be} \d_{cf} + (k_- + l_- + 2v_-) \d_{ab} \d_{cd} \d_{ef} - (a \leftrightarrow f)\Big ] \no
\\
&\qquad \qquad \qquad \qquad + (\text{cycle \ } k,c; \ l,d; \ v,e) \ .
\end{align}
The result for $A_{\rm exch}^{\rm (D)}$ is unambiguous as here the internal momentum is always off-shell:
\begin{align} \la{diff}
A_{\rm exch}^{\rm (D)} &= \tfrac{i}{144} \l^4\ r_+ \Big[ \frac{ k_- ^2 (-8 l_- + v_- ) + v_- (l_-^2 + 3 l_- v_- + 2 v_-^2) +
k_- (-8 l_-^2 - 14 l_- v_- + 3 v_-^2)}{ (k_- + v_- ) (l_- + v_- )} \d_{ab} \d_{cd} \d_{ef} \no \\
&\qquad \qquad\qquad + \frac{k_-^2 (l_- - 8 v_-) + 8 l_- v_- (l_- + v_-) - k_- (l_-^2 + 8 v_-^2) }{ (k_- + v_- ) (l_- + v_- )} \delta _{ad} \delta _{be} \delta_{cf} \\ &\qquad \qquad \qquad - (a \leftrightarrow f) \Big] + (\text{cycle \ } k,c; \ l,d; \ v,e) \ . \no
\end{align}
$A_{\rm exch}^{\rm (S)}$ has
Type 2 ambiguous contributions (as defined in \rf{2.20}) since here the
internal line is on-shell -- it may carry momenta $k$, $l$, $v$ or $k+l+v$
which have only ``$-$'' components. Let us focus, e.g., on the particular
diagram on the bottom left of Fig.\ref{6fig} with internal momentum $v_-$,
which we shall denote as $A_{\rm exch}^{\rm (S),v}$. Using the massive
regularization as defined in section \ref{kin} we are to consider the process
$X^a (z) X^b (w) \to X^c (k) X^d (l) X^e (v) X^f (r)$, where the momenta are
now on-shell with mass $m$, i.e. (cf. \rf{225})
\be \la{onshell}
z_- = -\tfrac{m^2}{z_+}\ , \ r_- = - \tfrac{m^2}{r_+} \ , \qquad \quad w_+ = -\tfrac{m^2}{w_-}\ , \ \ \ k_+ = -\tfrac{m^2}{k_-} \ ,\ \ \ l_+ = -\tfrac{m^2}{l_-}\ , \ \ \ v_+ = -\tfrac{m^2}{v_-} \ .
\ee
The momentum conservation conditions \rf{newCOM} are solved by
\begin{align} \la{COM}
&z_+ = r_+ + \xi m^2 + \O(m^5) \ , \quad \qquad
w_- = (k_- + l_- + v_-) - \xi \frac{m^4}{r_+^2} + \O(m^5) \ , \\
& \ \ \
\qquad \xi \equiv - \tfrac{(k_- + l_-) (k_- + v_-) (l_- + v_-) }{k_- l_- v_- (k_- + l_- + v_-)} \ .\no
\end{align}
Using the 4-vertex Feynman rule \rf{2.22} with ${\rm p}=1$, we obtain
\begin{align}
A_{\rm exch}^{\rm (S),v} = & - \tfrac{i}{144} \l^4\ \d_{ag}\d_{ef}\Big[ 2 z \cdot (v+r-z) + 2r \cdot v + (z+(v+r-z))\cdot (r+v)\Big] \\ \no
&\qquad \times \frac{1}{(v+r-z)^2+m^2}\ \d_{gb}\d_{cd}\Big[ 2 (z-r-v) \cdot w + 2k \cdot l + (z-r-v+w)\cdot (k+l)\Big] \\ \no
&\qquad \qquad \qquad + (\text{other tensor structures})
\end{align}
We presented here only the coefficient of $ \d_{ab} \d_{cd} \d_{ef} $ while the
coefficients of the other tensor structures are similar. Using \rf{onshell},
\rf{COM} we can write this diagram in terms of $r_+$, $k_-$, $l_-$, $v_-$ and
$m^2$ as
\begin{align}
&A_{\rm exch}^{\rm (S),v}
=- \tfrac{i}{144} \l^4\ \d_{ab} \d_{cd} \d_{ef} \ \ \big[ 3 r_+ v_- + \O(m^2)\big] \ \ \frac{1}{ \xi v_- m^2 + \O(m^4)}\no \\ \no
&\qquad\ \ \times \Big[ \Big( \xi (k_- + l_- + v_-) + \tfrac{k_- + l_- + v_-}{v_-} + \tfrac{v_-}{k_- + l_- + v_-} - \tfrac{k_-}{l_-} - \tfrac{l_-}{k_-} \\ \no
&\qquad \qquad + \ha(\xi + \tfrac{1}{v_-} - \tfrac{1}{k_- + l_- + v_-})(k_- + l_-) + \ha (v_- - (k_- + l_- + v_-)(\tfrac{1}{k_-} + \tfrac{1}{l_-}) \Big) m^2 + \O(m^4) \Big] \\
&\qquad \qquad \qquad \qquad + (\text{other tensor structures}) \ . \la{3.7}
\end{align}
Here the first vertex factor $[3 r_+ v_- + \O(m^2)]$ is finite as $m\to 0$,
while the propagator and the second vertex are of order $\tfrac{1}{m^2}$ and
$m^2$ respectively, so that their product has a finite massless limit:
\begin{align}
A_{\rm exch}^{\rm (S),v} &=
\tfrac{i}{48} \l^4 \d_{ab} \d_{cd} \d_{ef} \frac{r_+ v_- \big[ 3 k_-^3 + k_-^2 (6 l_- + 4 v_-) +
k_- (6 l_-^2 + 4 l_- v_- + v_-^2) + l_- (3 l_-^2 + 4 l_- v_- + v_-^2)\big] }{(k_- + l_-) (k_- + v_-) (l_- + v_-)} \no\ \ \\
& \qquad \qquad \qquad \qquad + (\text{other tensor structures})\ . \la{3.8}
\end{align}
Summing up all similar contributions gives
\begin{align} \la{same}
A_{\rm exch}^{\rm (S)} &= A_{\rm exch}^{\rm (S),v} + A_{\rm exch}^{\rm (S),k} + A_{\rm exch}^{\rm (S),l} + A_{\rm exch}^{\rm (S),k+l+v} \\ \no
&= - \tfrac{i}{48} \l^4 r_+ \big[(k_- - l_-) \d_{ad} \d_{be} \d_{cf} + (k_- + l_- + 2v_-) \d_{ab} \d_{cd} \d_{ef} - (a \leftrightarrow f) \big] + (\text{cycle } k,c; l,d; v,e)
\end{align}
Finally, adding together \rf{cont}, \rf{diff}, \rf{same}, we find for the
amplitude \rf{3.1}
\begin{align}
&S[X^a (r_+) X^b (k_- +l_- +v_-) \to X^c (k_-) X^d (l_-) X^e (v_-) X^f (r_+)] \la{3.10} \\
&\qquad = \tfrac{i}{16} \l^4\ r_+ \Big[- \frac{ k_- l_- (k_- + l_- +2 v_- )}{(k_- +v_- ) (l_- +v_- )} \delta _{ab} \delta _{cd} \delta _{ef} + \frac{ v_- (l_- -k_- ) (k_- +l_- +v_- )}{(k_- +v_- ) (l_- +v_- )} \delta _{ad} \delta _{be} \delta_{cf} \no \\
& \qquad \qquad \qquad\ \ \ \ \ \ - (a \leftrightarrow f) \Big ]
+ (\text{cycle } \ k,c; \ l,d; \ v,e) \ . \no
\end{align}
This amplitude is non-vanishing for a generic set of minus-momenta and does not factorize. %v3

Similarly, for the particle-production
$+- \to --++$ amplitude we get the following non-zero result
\begin{align}
&S[X^a (v_+ + r_+) X^b (k_- + l_-) \to X^c (k_-) X^d (l_-) X^e (v_+) X^f (r_+)] \la{244} \\
&\qquad = - \tfrac{i}{16} \l^4\ \Big[ v_+ k_- \d_{af} \d_{bd} \d_{ce} + (\text{cycle } \ k,c; \ l,d; \ -(k+l),b ) \Big]
+ (\text{cycle} \ v,e; \ r,f; \ -(r+v),a ) \ . \no
\end{align}

Analogous results can be obtained for coset $\s$-models. The $SU(2)$ PCM has
target space $SU(2) \cong S^3 \cong SO(4)/ SO(3)$, and indeed the amplitudes
\rf{3.10}, \rf{244} turn out to be identical to the corresponding amplitudes in
the \sm on $SO(4)/SO(3)$.\foot{While this model has non-perturbative $SU(2)
\times SU(2)$ or $SO(4)$ symmetry, this symmetry is broken to $SO(3)$ in the
perturbative expansion near the trivial vacuum point.}
The $SU(2)$ PCM and %v3
$SO(4)/SO(3)$ models may be described using different natural choices of
coordinates, i.e. related by an $SO(3)$-symmetric field redefinition. In Appendix \ref{equiv}, we show explicitly that the amplitudes \rf{3.10}, \rf{244} are invariant
under all such $SO(3)$-symmetric redefinitions (provided one uses the massive regularization), while the S-matrix
equivalence theorem is generally anomalous for non-symmetric redefinitions.

The amplitudes \rf{3.10}, \rf{244} have exactly the same form for any $S^N =
SO(N+1)/SO(N)$ model\foot{A similar non-vanishing expression for the amplitude
\rf{3.10} was found in \cite{Figueirido:1988ct} and we confirm the conclusion
of \cite{Figueirido:1988ct} about the lack of factorization of the tree level
S-matrix in the $S^N$ model. } written in the embedding coordinates ($(X^a)^2
+ (X^{N+1})^2 = 4 \l^{-2}, \ a=1,\dots,N$)
\be\la{2445}
\L = -\ha \left[ (\del X^a)^2 + ( \del X^{N+1} )^2 \right]
= -\ha \Big[ (\del X^a)^2 + \frac{\tfrac{\l^2}{4} (X^a \del X^a)^2 }{1-\tfrac{\l^2}{4} (X^a)^2 } \Big] \ .
\ee
The alternative $SO(N)$-symmetric coordinates \rf{eq:onsm} for $S^N$, %v3
considered in Appendix \ref{comments}, are again related by an $SO(N)$-symmetric redefinition, and thus one similarly obtains exactly the same
mass-regularized amplitudes \rf{3.10}, \rf{244}.\foot{Note that the coordinates \rf{eq:onsm} generalize straightforwardly to $SO(N+1)/SO(N)$ for general $N$, though they are considered specifically in the $N=3$ case in Appendix C.}

The amplitudes \rf{3.10}, \rf{244} are non-zero as the coefficients of
independent $SO(N)$ tensor structures are non-zero.
For example, for
$(a,b,c,d,e,f) = (1,1,2,2,2,2)$ we get \begin{align} &S[X^1 (r_+) X^1 (k_- +l_-
+v_-) \to X^2 (k_-) X^2 (l_-) X^2 (v_-) X^2 (r_+)] \no \\ &\qquad \qquad
= - \tfrac{i}{16} \l^4\ r_+ (k_- + l_- + v_-) \ \la{246},\\
&S[X^1 (r_+ + v_+) X^1 (k_- + l_-) \to X^2 (k_-) X^2 (l_-) X^2 (v_+) X^2 (r_+)] \no \\&\qquad \qquad
= - \tfrac{i}{16} \l^4\ (r_+ + v_+)( k_- + l_-) \ . \la{247}
\end{align}
Thus, despite being classically integrable, the PCM and $SO(N+1)/SO(N)$ coset %v3
\sms have tree-level 6-point amplitudes that either have massless particle production
or do not factorize.

%%%%%%%%%%%%%%%%%%%%%%%%%%%%%%%%%%%%%%
\subsection{\texorpdfstring{$2 \to 3$}{2 to 3} amplitude in the \texorpdfstring{ZM$_q$}{ZMq} model}\label{ZMamp}
%%%%%%%%%%%%%%%%%%%%%%%%%%%%%%%%%%%%%%

Next, let us consider the 5-point amplitude $X^b (r_+) X^c (k_- + l_-) \to X^a
(r_+) X^d (k_-) X^e (l_-) $ in the ZM$_q$ model that was found to be non-zero
in \cite{Nappi:1979ig} (in the $q=0$ case). We will confirm this in both the
massive regularization and the $i\e$-regularization.

The Feynman rules for the ZM$_q$ model are given by \rf{2.21}, \rf{2.23} with
$\qq=1-q^2$ (see Table 1). The 5-point amplitude is built out of the exchange
diagrams with three 3-vertices and two internal propagators. Assuming that the
three outgoing particles have the same labels $a=d=e$, and using that the
3-vertex \rf{2.23} is proportional to $f_{abc}$ and hence totally
anti-symmetric, the only contributions are the 6 diagrams (corresponding to the
different permutations of the three $X^a$ legs) with $X^b$ and $X^c$ incident
to different vertices. We group them into pairs $A^{(1)}$, $A^{(2)}$,
$A^{(3)}$ of diagrams related by swapping the $k_-$ and $l_-$ legs (one of each
pair is shown in Fig.\ref{ZMfig}):
\begin{figure}[t!]
\centering
\raisebox{-0.5\height}{\includegraphics[scale=1.05]{\figpath/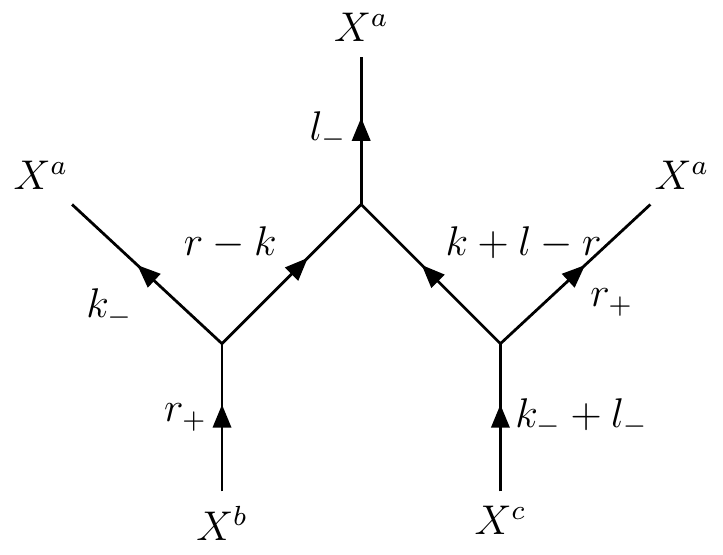}} \hspace{1.8cm} \raisebox{-0.5\height}{\includegraphics[scale=1.05]{\figpath/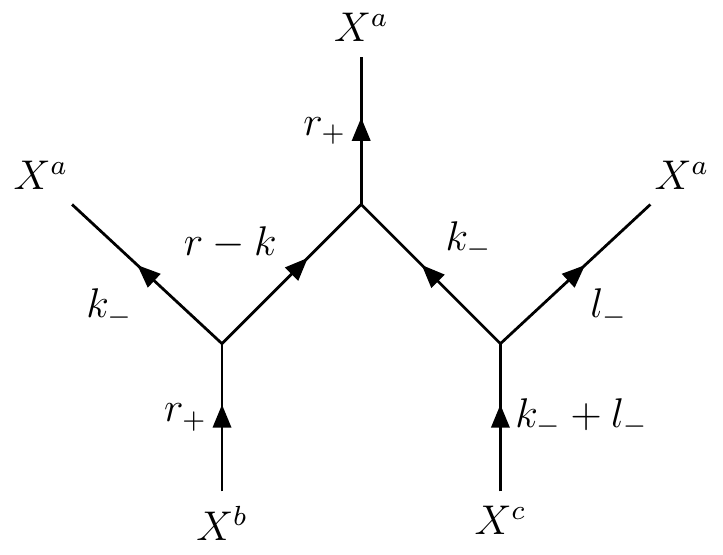}} \\
\raisebox{-0.5\height}{\includegraphics[scale=1.05]{\figpath/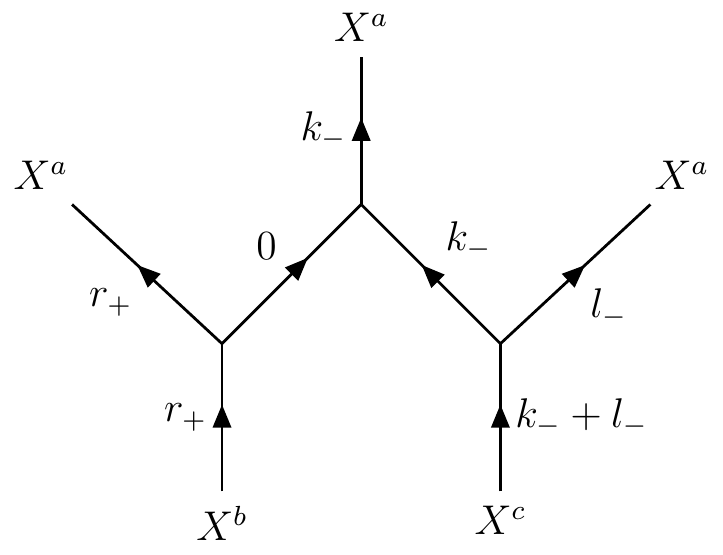}}
\caption{\small Contributions to the 2 $\to$ 3 amplitude \rf{ZMpp}. Left: Unambiguous diagram $A^{(1)}$. Right: Diagram $A^{(2)}$ with a Type 2 ambiguity.
Bottom: Diagram $A^{(3)}$ with Type 1 and Type 2 ambiguities. Each diagram is accompanied by the corresponding diagram with the $k_-$ and $l_-$ legs swapped. \label{ZMfig}}
\end{figure}
\be \la{2.31}
S[X^b (r_+) X^c (k_- + l_-) \to X^a (r_+) X^a (k_-) X^a (l_-)] = A^{(1)} +A^{(2)} +A^{(3)} \ . \ee
$A^{(1)}$ is given by an unambiguous expression (with no summation over fixed index $a$):
\begin{align}
A^{(1)}
&= - \tfrac{i}{128} \l^3 \ (1-q^2)^3 \ f^{abd} {f^{ae}}_d {{f^a}_e}^c \ \ r_+ (k_- +l_-) \ . \la{2.32}
\end{align}
The contributions $A^{(2)}$ and $A^{(3)}$ are ambiguous. To define them let us first use the massive regularization. Analogously to the 6-point case discussed in section \ref{6point}, we consider the massive process $X^b (z) X^c (w) \to X^a (r) X^d (k) X^e (l)$ with momenta
\be \la{2to3m}
z_- = -\tfrac{m^2}{z_+}\ , \ r_- = - \tfrac{m^2}{r_+} \ , \qquad \qquad \quad w_+ = -\tfrac{m^2}{w_-}\ , \ \ \ k_+ = -\tfrac{m^2}{k_-} \ ,\ \ \ l_+ = -\tfrac{m^2}{l_-} \ .
\ee
of mass $m$, whose limits give the desired massless momenta as $m\to 0$. We solve conservation of momentum \rf{newCOM} by
\begin{align} \la{2to3com}
&z_+ = r_+ + \xi m^2 + \O(m^5) \ , \quad \qquad
w_- = (k_- + l_- ) - \xi \frac{m^4}{r_+^2} + \O(m^5) \ , \\
& \ \ \
\qquad \xi \equiv - \tfrac{k_-^2 + k_- l_- + l_-^2 }{k_- l_- (k_- + l_- )} \ .\no
\end{align}
The ambiguity in $A^{(2)}$ is only of Type 2, since the internal momenta $k$ and $l$ are on-shell but non-vanishing, and thus it is expected to be non-zero. Indeed, starting with the massive momentum configuration \rf{2to3m}, \rf{2to3com}, we find in the $m \to 0$ limit
\begin{align}
A^{(2)} = A^{(1)} + \O(m^2) \to A^{(1)} = - \tfrac{i}{128} \l^3(1-q^2)^3 \ f^{abd} {f^{ae}}_d {{f^a}_e}^c \ \ r_+ (k_- +l_-) \ .
\end{align}
$A^{(3)}$ contains both Type 1 and Type 2 ambiguities. As discussed in section \ref{kin}, Type 1 ambiguities are vanishing in the massive regularization, and so we find that in the $m \to 0$ limit
\begin{align}
A^{(3)} \propto (\text{Type 1}) \times (\text{Type 2}) = \O(m^2) \times \O(m^0) \to 0\ .
\end{align}
Hence the total amplitude in the massive regularization is\foot{In the $i\e$-regularization that
was seemingly used in \cite{Nappi:1979ig}, all ambiguous contributions are instead resolved as zero; in particular,
$A^{(2)}=0$ and the amplitude $S=A^{(1)}$ is half that of the result \rf{ZMpp} found using the massive regularization.}
\begin{align}\la{ZMpp}
&S[X^b (r_+) X^c (k_- + l_-) \to X^a (r_+) X^a (k_-) X^a (l_-)] = 2 A^{(1)} \no\\
&\qquad \qquad \qquad \qquad \qquad \qquad \qquad = - \tfrac{i}{64} \l^3 \ (1-q^2)^3 \ f^{abd} {f^{ae}}_d {{f^a}_e}^c \ r_+ (k_- + l_-) \ .
\end{align}
For example, in the $SU(2)$ case where $ f^{abd} {f^{ae}}_d {{f^a}_e}^c = 16 \e^{abc}$, it is
\be \la{ZMsu2}
S[X^b (r_+) X^c (k_- + l_-) \to X^a (r_+) X^a (k_-) X^a (l_-)] = - \tfrac{i}{4} \l^3 \ (1-q^2)^3 \ \e^{abc} r_+ (k_- + l_-) \ .
\ee
This amplitude is non-zero for all values of $q$ except $q=\pm 1$, where the theory is free.
Thus, despite being classically integrable, the ZM$_q$ model
exhibits tree-level massless particle production.
\begin{figure}[t!]
\centering
\raisebox{-0.5\height}{\includegraphics[scale=0.95]{\figpath/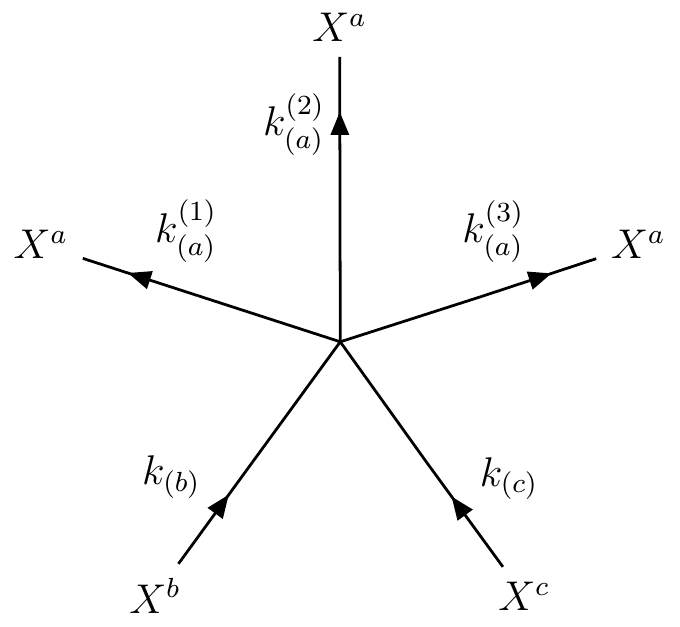}}\hspace{2cm} \raisebox{-0.5\height}{\includegraphics[scale=0.95]{\figpath/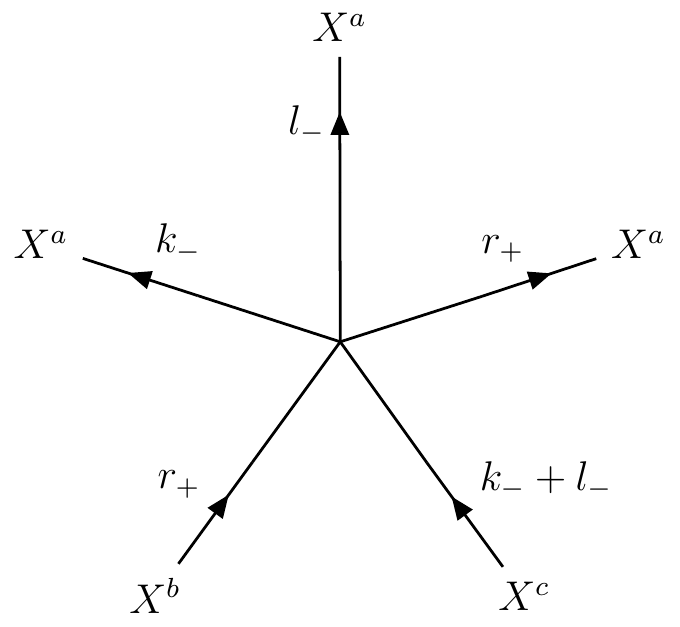}} \\
\vspace{0.3cm}
\raisebox{-0.5\height}{\includegraphics[scale=1.05]{\figpath/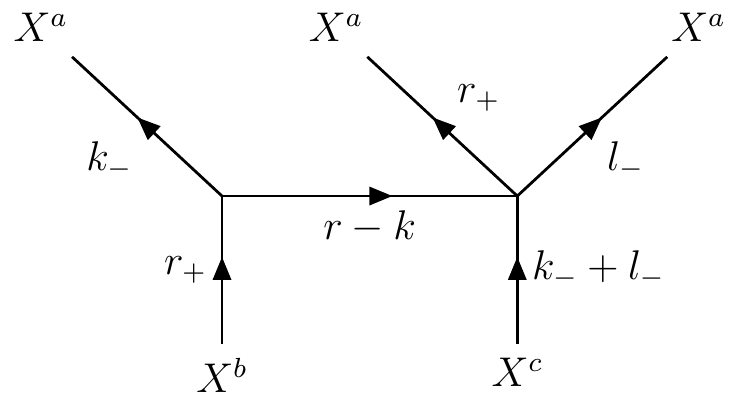}} \hspace{1.3cm} \raisebox{-0.5\height}{\includegraphics[scale=1.05]{\figpath/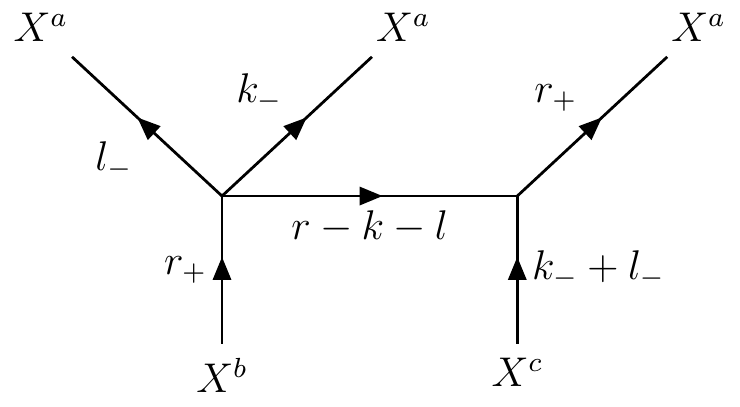}}\\
\raisebox{-0.5\height}{\includegraphics[scale=1.05]{\figpath/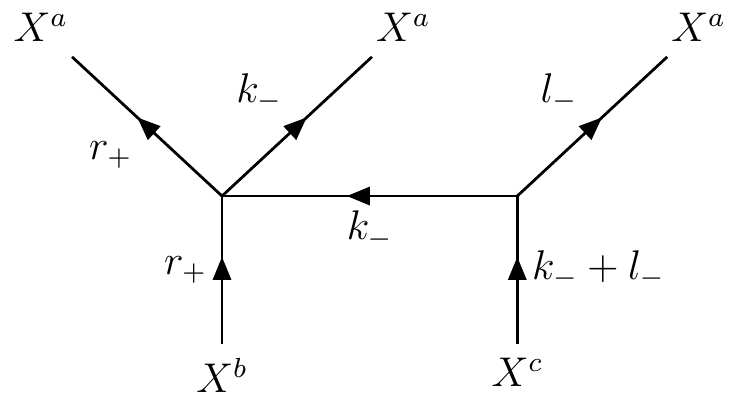}} \hspace{1.3cm} \raisebox{-0.5\height}{\includegraphics[scale=1.05]{\figpath/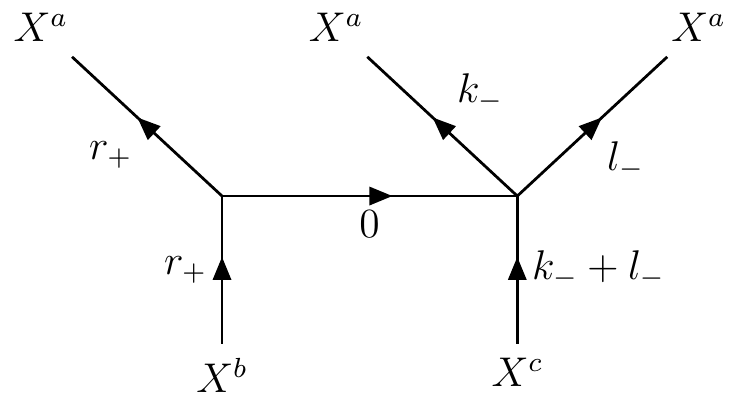}}
\caption{\small Diagrams contributing to the 2 $\to$ 3 amplitude \rf{NADpp}, in addition to those in Fig.\ref{ZMfig}. Top left: Off-shell 5-vertex $V_{bcaaa}$.
Top right: On-shell contact diagram $A_{\rm cont}$.
Middle row: unambiguous exchange diagrams $A_{\rm unambig}$.
Bottom row: Ambiguous exchange diagrams $A_{\rm ambig}$. The left-hand exchange diagrams are accompanied by the corresponding diagrams with the $k$ and $l$ legs swapped. \label{NADfig}}
\end{figure}

%%%%%%%%%%%%%%%%%%%%%%%%%%%%%%%%%%%%%%
\subsection{\texorpdfstring{$2 \to 3$}{2 to 3} amplitude in the NAD model}
%%%%%%%%%%%%%%%%%%%%%%%%%%%%%%%%%%%%%%

Now let us compute the same amplitude $X^b (r_+) X^c (k_- + l_-) \to X^a (r_+)
X^d (k_-) X^e (l_-)$ in the non-abelian dual model \rf{2.16} (specializing to
the $SU(2)$ case $\g_{ab}=-\ha \d_{ab} , \ f_{abc} = -\ha \e_{abc} , \
a,b,c=1,2,3$). We will use the massive regularization to resolve the
ambiguities, and will again assume that the three outgoing particles have the
same labels $a=d=e$.

The 3- and 4-vertex Feynman rules are \rf{2.20}, \rf{2.21}, \rf{2.22} with
$\pp=12$, $\qq=-3$ (see Table 1). Since the diagrams in Fig.\ref{ZMfig}
contributing to the ZM ($q=0$) amplitude \rf{ZMsu2} only contain 3-vertices,
the NAD amplitude gets contributions from all of these diagrams. The 3-vertices
are related by $V_{(3)}^{\rm NAD} = -3 V_{(3)}^{{\rm ZM}}$ and as the ZM
amplitude is cubic in the 3-vertex, the corresponding contribution to the
NAD amplitude is
\be \la{NAD3vert}
A_{\text{3-vertices}} = (-3)^3 S^{{\rm ZM}}
= \tfrac{27}{4} \l^3 \e^{abc} r_+ (k_- + l_-) \ .
\ee
The 5-point term in the Lagrangian \rf{2.16} is (here the index $d$ is contracted with $\d_{ab}$)
\be \la{NAD5pt}
\mathcal{L}^{(5)}_{\rm NAD} = -\tfrac{\l^3}{2} \e^{\m\n} \e_{abc} X^d X^d X^a \del_\m X^b \del_\n X^c \ .
\ee
Thus the Feynman rule for the 5-vertex shown in the top left of Fig.\ref{NADfig} is (for $a=d=e$)
\be
V_{bcaaa} = 10 i \l^3 \e_{abc} \e_{\m\n} k_{(b)}^\m k_{(c)}^\n \ .
\ee
Putting this 5-vertex on-shell gives the contact diagram in the top right of Fig.\ref{NADfig}:\foot{Here $\e_{\m\n} r^\m (k+l)^\n = \ha r_+ (k_- + l_-)$.}
\be \la{NADcont}
A_{\rm cont} = 5 i \l^3 \e_{abc} r_+ (k_- + l_-) \ .
\ee
The remaining diagrams in Fig.\ref{NADfig} are exchanges with one 3-vertex and one 4-vertex. Two of these are unambiguous, contributing
\be \la{NADunambig}
A_{\rm unambig} = (-6-\tfrac{3}{2})\ i \l^3 r_+ (k_- + l_-) = -\tfrac{15}{2}\, i \l^3 r_+ (k_- + l_-) \ .
\ee
One of the ambiguous diagrams has a Type 1 ambiguity and so vanishes in the massive regularization. The other has only a Type 2 ambiguity and, as expected, is
non-zero in the massive regularization
\be \la{NADambig}
A_{\rm ambig} = -\tfrac{9}{2} i \l^3 r_+ (k_- + l_-)\ .
\ee
Summing up the contributions \rf{NAD3vert}, \rf{NADcont}, \rf{NADunambig}, \rf{NADambig} we find
\begin{align}
&S[X^b (r_+) X^c (k_- + l_-) \to X^a (r_+) X^a (k_-) X^a (l_-)] = A_{\text{3-vertices}} + A_{\rm cont} + A_{\rm unambig} + A_{\rm ambig} \no \\
&\qquad \qquad = ( \tfrac{27}{4} + 5 - \tfrac{15}{2} - \tfrac{9}{2})\ i \l^3 \e_{abc} r_+ (k_- + l_-) = - \tfrac{i}{4} \, \l^3 \e_{abc} r_+ (k_- + l_-) \ . \la{NADpp}
\end{align}
Curiously, the amplitude \rf{NADpp} is equal to the corresponding particle-production amplitude \rf{ZMpp} in the ZM model (with $q=0$),
although the significance of this fact is not clear.

In section \ref{4-pt} we saw that the 4-point amplitudes in PCM and NAD
differ by an overall factor of 3
(see Table 2). The two models differ even more drastically at the 5-point level:
while the PCM has only even vertices so has vanishing 5-point amplitude, the NAD has non-zero 5-point particle-production amplitudes. This demonstrates again that, contrary to naive expectations,
the arguments about path integral duality between the PCM and NAD
models \cite{Fridling:1983ha,ftd} do not imply the equality of the corresponding massless S-matrices.

%%%%%%%%%%%%%%%%%%%%%%%%%%%%%%%%%%%%%%
\subsection{\texorpdfstring{$2 \to 3$}{2 to 3} amplitude in the \texorpdfstring{PCM$_q$}{PCMq}}
%%%%%%%%%%%%%%%%%%%%%%%%%%%%%%%%%%%%%%

Finally, let us compute the same 5-point
amplitude $X^b (r_+) X^c (k_- + l_-) \to X^a (r_+) X^d (k_-) X^e (l_-) $ (with $a=d=e$) in PCM$_q$ with non-zero coefficient $q$ of the WZ term.
The structure of the computation is exactly the same (with the same diagrams)
as in the NAD model
since the $n$-point vertices $V_{(n)}$ are the same up to
numerical factors:\foot{The factor $-\tfrac{q}{60}$ for the 5-vertex follows from
comparing the 5-point term in the PCM$_q$ Lagrangian \rf{pcmq}, $\mathcal{L}_{{\rm PCM}_q}^{(5)} = \tfrac{\l^3 q}{120} \e^{\m\n} \e_{abc} X^d X^d X^a \del_\m X^b \del_\n X^c ,$ with \rf{NAD5pt}.}
\be
V_{(3)}^{{\rm PCM}_q} = -\tfrac{q}{3} V_{(3)}^{\rm NAD} \ , \qquad V_{(4)}^{{\rm PCM}_q} = \tfrac{1}{12} V_{(4)}^{\rm NAD} \ , \qquad V_{(5)}^{{\rm PCM}_q} =- \tfrac{q}{60} V_{(5)}^{\rm NAD} \ .
\ee
Then using the expressions \rf{NADcont}, \rf{NAD3vert}, \rf{NADunambig}, \rf{NADambig}
found in the NAD case above, one finds for the corresponding diagrams in PCM$_q$:
\begin{align}
A_{\rm cont}^{{\rm PCM}_q} &=- \tfrac{q}{60} A_{\rm cont}^{\rm NAD} =- \tfrac{i}{12} \l^3 \, q\, \e_{abc} r_+ (k_- + l_-)\ , \\
A_{\text{3-vertices}}^{{\rm PCM}_q} &= ( -\tfrac{q}{3} )^3 A_{\text{3-vertices}}^{\rm NAD} = - \tfrac{i}{4} \l^3\, q^3\, \e_{abc} r_+ (k_- + l_-)\ , \\
A_{\rm unambig}^{{\rm PCM}_q} &= (-\tfrac{q}{3})(\tfrac{1}{12}) A_{\rm unambig}^{\rm NAD} = \tfrac{5i}{24} \l^3\, q\, \e_{abc} r_+ (k_- + l_-)\ , \\
A_{\rm ambig}^{{\rm PCM}_q} &= (-\tfrac{q}{3})(\tfrac{1}{12}) A_{\rm ambig}^{\rm NAD} = \tfrac{i}{8} \l^3\, q\, \e_{abc} r_+ (k_- + l_-) \ .
\end{align}
These contributions sum up to (cf. \rf{NADpp})\foot{The expression \rf{PCMqpp}, which was found in the massive regularization, would be multiplied by an extra factor of $\ha$ if computed in
the $i \e$-regularization (used in \cite{Figueirido:1988ct}). To see this recall from section \ref{ZMamp} that in the $i\e$-regularization the $A_{\text{3-vertices}}$ contribution would halve and the ambiguous term $A_{\rm ambig}$ would
be set to zero. }
\be \la{PCMqpp}
S[X^b (r_+) X^c (k_- + l_-) \to X^a (r_+) X^a (k_-) X^a (l_-)] = -\tfrac{i}{4} q (q^2 - 1) \l^3 \e_{abc} r_+ (k_- + l_-) \ .
\ee
For general $q$ this is non-zero, and thus there is
particle production in PCM$_q$, already at the 5-point level.
The amplitude \rf{PCMqpp} vanishes at the WZW
points $q=\pm 1$, complementing
the vanishing of the 4-point amplitudes in the WZW model observed in
section \ref{4-pt} and confirming
that the massless S-matrix of the WZW
model should be trivial
due to the decoupling of the
left-moving and right-moving modes.
It also vanishes in the PCM case with $q=0$ where all the vertices are even.

%%%%%%%%%%%%%%%%%%%%%%%%%%%%%%%%%%%%%%
\section{Massless scattering in doubled formalism and T-duality} \la{Tduality}
%%%%%%%%%%%%%%%%%%%%%%%%%%%%%%%%%%%%%%

Given that there are IR ambiguities in the scattering amplitudes of 2d chiral scalars computed in the standard way
one may wonder if a better definition of the massless S-matrix
may be achieved in the ``doubled'' formulation \cite{rt} (see Appendix \ref{doubled}).
The idea is to treat the left and right chiral scalars as independent
off-shell fields (at the expense of off-shell 2d Lorentz invariance).
The resulting S-matrix is then automatically duality-invariant, and retains on-shell Lorentz invariance.

Expanding the metric as $G_{ab} (X) = \delta_{ab} + V_{ab} (X)$,
the doubled Lagrangian \rf{4} may be written as (here we set $X^a \equiv x^a$ and we use \rf{5}, \rf{6})
\begin{align}
\L = &- \del_1 X_+^a \del_- X_+^a - \del_1 X_-^a \del_+ X_-^a \no \\
&- V_{ab}(X) \del_1 X_+^a \del_1 X_-^b - B_{ab}(X) \del_1 X_+^a \del_1 X_-^b + \O(B^2) + \O(V^2) + \O(VB) \ . \la{4.1}
\end{align}
At linear order in $V$ and $B$ there are no ``chiral'' vertices involving only $X_+$ or only $X_-$.
The $X_\pm$ particles have respective propagators $\Delta_\pm (k) = \tfrac{-i}{2 k_1 k_\mp}$.
As a result, one can see that, with only the lowest-order vertices linear in $V$ and $B$,
there will be no Type 1 or Type 2 ambiguities in simple exchange diagrams with just one internal line.
At higher orders there may still be ambiguities, which may be resolved as discussed in sections \ref{tree} and \ref{higher}.
In this section we will focus on such simple low-order amplitudes that are protected from
ambiguities and are thus naturally well defined in the doubled formulation.

One property of the doubled \sm discussed in Appendix \ref{doubled}
is that T-duality becomes a manifest symmetry.
Let us check directly that the S-matrix computed in the doubled formulation
is indeed T-duality covariant on a simple example.
We shall consider the following \sm Lagrangian $\L$ with an abelian isometry in the $Y$-direction,
and its T-dual $\tilde{\L}$
\be\la{4.2}
\L = -\ha (\del X)^2 -\ha (1+ \l^2 X^2) (\del Y)^2 \ , \qquad \qquad \tilde{\L} = -\ha (\del X)^2 -\ha ({1+ \l^2 X^2})^{-1} (\del Y)^2 \ .
\ee
Here we formally denote the isometric field of both the original and dual theories by $Y$.
The respective doubled Lagrangians \rf{4} for the two models in \rf{4.2},
\begin{align}
\L_{\rm double} &= \L_0 - \ha (1+\l^2 X^2) (\del_1 Y)^2 - \ha ({1+ \l^2 X^2})^{-1}(\del_1 \tilde{Y})^2 \ ,\la{doub1}\\
\tilde{\L}_{\rm double} &= \L_0 - \ha ({1+ \l^2 X^2})^{-1} (\del_1 Y)^2 - \ha (1+ \l^2 X^2)(\del_1 \tilde{Y})^2 \ ,\la{doub2}\\
\L_0 &\equiv \ha \big( \del_0 X \del_1 \tilde{X} + \del_1 X \del_0 \tilde{X} + \del_0 Y \del_1 \tilde{Y} + \del_1 Y \del_0 \tilde{Y} \big) \ ,
\end{align}
are equivalent, being related by the transformation
$Y \leftrightarrow \tilde{Y}$.
This is a special case of \rf{44}, \rf{55} with $x\to Y, z\to X$.
Written in the chiral basis $Y_\pm = \ha (Y \pm \tilde{Y})$
this transformation is equivalent to flipping the sign of $Y_-$ and thus
the scattering amplitudes corresponding to \rf{doub1} and \rf{doub2} should be related as \cite{rt}
\be \la{4.3} \tilde{S} = (-1)^{n_-} \ S \ , \ \ \ \ \ \ \ \ \ \ \ \
Y_+ \to Y_+ \ , \qquad Y_- \to -Y_- \ , \ee
where $n_-$ is the number of $Y_-$ fields being scattered.
If we restrict our attention to 6-point amplitudes, then
such doubled amplitudes will have no ambiguities in view of the above discussion
(exchange diagrams will contain only a single internal line and only non-chiral vertices linear in $V$ and $B$ in \rf{4.1}).
The only non-vanishing 6-point amplitudes for $\L_{\rm double}$ in \rf{doub1} are found to be\foot{Here
$Y(k_+)\equiv Y_+ (k_+), \ Y(k_-)\equiv Y_- (k_-) $, etc.
As the free fields satisfy $\del_- Y_+=0, \ \del_+ Y_-=0$ the non-zero momenta of
the on-shell left-moving $Y_+ (\sigma^+) $ and right-moving $Y_-(\sigma^-) $ are, respectively, $k_+$ and $k_-$.}
\begin{align}
&S[Y (r_+) Y (k_- +l_- +v_-) \to X(k_-) X (l_-) X (v_-) X (r_+)] = -6i \l^4\, r_+ (k_- + l_- + v_-)\ ,\la{4.7} \\
&S[Y (v_+ + r_+) Y (k_- + l_-) \to X (k_-) X (l_-) X (v_+) X (r_+)] = -6i \l^4\, (v_+ + r_+) (k_- + l_-) \ , \la{4.8}
\end{align}
plus those related to \rf{4.7}, \rf{4.8} by crossing symmetry.
The corresponding amplitudes for $\tilde{\L}_{\rm double}$ in \rf{doub2} differ by a sign
\begin{align}
&\tilde{S}[Y (r_+) Y (k_- +l_- +v_-) \to X(k_-) X (l_-) X (v_-) X (r_+)] = +6i \l^4\, r_+ (k_- + l_- + v_-)\ ,\\
&\tilde{S}[Y (v_+ + r_+) Y (k_- + l_-) \to X (k_-) X (l_-) X (v_+) X (r_+)] = +6i \l^4\, (v_+ + r_+) (k_- + l_-) \ ,
\end{align}
as expected, since the number of the $Y_-$ fields being scattered is $n_-=1$.

Next, let us compute some of the amplitudes in sections \ref{tree} and \ref{higher}, this time using the doubled formulation
of the corresponding $\s$-models.
Let us start with the 4-point amplitude for the ``interpolating'' Lagrangian $\L_{\pp, \qq}$ in \rf{2.17}
in the $SU(2)$ case. Since any exchange diagram here is simple and contains only the lowest order cubic vertices (cf. Fig.\ref{fig2}) the
corresponding amplitude is unambiguous and we get
\be
S[X^a (k_+) X^b (l_-) \rightarrow X^c (k_+) X^d (l_-)] =
- \tfrac{i}{4} \l^2 (\pp - \tfrac{13}{9} \qq^2) (\d_{ab}\d_{cd} - \d_{ad}\d_{cb})\ k_+ l_- \ . \label{4double}
\ee
Note that the $\pp$-dependence matches the previous result in \rf{2.30} but the $\qq$-dependence does not.

The $2 \to 4$ amplitudes in the $SU(2)$ PCM and the $S^N = SO(N+1)/SO(N)$ coset model were
computed in section \ref{6point} in the massive regularization.
As there is no $B$-field, here the lowest order vertex is quartic; thus the exchange diagrams
are simple and contain only quartic vertices (cf. Fig.\ref{6fig}), so the resulting amplitude in the doubled formulation
is again unambiguous\foot{Let us note that the same results are obtained in all three of the following of coordinate choices on $S^N$: \rf{2.1}-\rf{2.3}, \rf{2445} (with $N=3$) and \rf{eq:onsm}.}
\begin{align}
&S[X^a (r_+) X^b (k_- +l_- +v_-) \to X^c (k_-) X^d (l_-) X^e (v_-) X^f (r_+)] \la{16} \\
&\qquad = \tfrac{i}{16} \l^4\ r_+ \Big[- \frac{ k_- l_- (k_- + l_- +2 v_- )}{(k_- +v_- ) (l_- +v_- )} \delta _{ab} \delta _{cd} \delta _{ef} + \frac{ v_- (l_- -k_- ) (k_- +l_- +v_- )}{(k_- +v_- ) (l_- +v_- )} \delta _{ad} \delta _{be} \delta_{cf} \no \\
& \qquad \qquad \qquad\ \ \ \ \ \ - (a \leftrightarrow f) \Big ]
\quad + (\text{cycle } \ k,c; \ l,d; \ v,e) \ , \no \\
&S[X^a (v_+ + r_+) X^b (k_- + l_-) \to X^c (k_-) X^d (l_-) X^e (v_+) X^f (r_+)] \la{17} \\
&\qquad = - \tfrac{5i}{64} \l^4\ \Big[ v_+ k_- \d_{af} \d_{bd} \d_{ce} + (\text{cycle } \ k,c; \ l,d; \ -(k+l),b ) \Big]
+ (\text{cycle} \ v,e; \ r,f; \ -(r+v),a ) \ . \no
\end{align}
The first amplitude \rf{16} matches the previous result \rf{3.10}.
The second amplitude \rf{17} has the same form as \rf{244} except with coefficient $-\tfrac{5i}{64}$ instead of $-\tfrac{i}{16}$.

We conclude that the amplitudes found in the doubled formalism and the ones
found in the standard approach using the massive regularization do not always
match. The examples where they differ (\rf{17} and the $\qq$-dependence in
\rf{4double}) are precisely those where the amplitudes computed in the standard
approach feature Type 1 ambiguities, i.e. get contributions from diagrams with
vanishing internal momenta. The reason for this disagreement should be related
to the non-local field-dependent nature of the transformation between the
fields in the standard and the doubled action and thus, effectively, to the
different ways of how the IR ambiguities appear and are resolved in the two
approaches.

Closely related is the observation that, while the S-matrices of two T-dual
\sms like \rf{4.2} computed in the doubled approach are equivalent (mapped to
each other according to \rf{4.3}), this is not so in general in the standard
approach using the massive or $i\e$-regularization for Type 1 ambiguities
(which amounts to setting them to zero). As already discussed in the context
of the NAD model in section \ref{4-pt}, this may be attributed to the fact that
the relation between the original and dual fields is non-local (for example, in
models like \rf{4.2} we get $\del_a Y \to \epsilon_{ab} [G (X)]^{-1} \del^b \td
Y$).

%%%%%%%%%%%%%%%%%%%%%%%%%%%%%%%%%%%%%%
\section{Concluding remarks}
%%%%%%%%%%%%%%%%%%%%%%%%%%%%%%%%%%%%%%

In this paper we computed the tree-level massless S-matrices of the PCM and
related models, emphasizing the issue of on-shell IR ambiguities in scattering
of chiral 2d scalars. We found that the 4-point amplitudes of duality-related
models generally differ by an overall constant factor, while they all take the
same universal form due to the group symmetry. The fact that these amplitudes
do not coincide should be due to non-locality and non-linearity of the duality
relations between the corresponding fields.

At 5- and 6-points, we found that these classically integrable models have
non-zero particle production amplitudes implying that the usual association
between integrability and absence of particle production does not directly
apply in the massless scalar scattering case. This may suggest reconsidering the
approach used in \cite{Wulff:2017vhv}, where the presence of particle
production in some massless amplitudes was used to determine the
non-integrability of certain additional $B$-field couplings in some symmetric
space $\s$-models.

One may attempt to attribute the particle production to the presence of IR
ambiguities in these amplitudes, whose regularization (e.g., by a small mass
parameter) effectively breaks the integrability of the theory. It would be
interesting to check if this is indeed the case, i.e. if all massless
particle-production amplitudes that are free from IR ambiguities vanish in
integrable models. This requires further investigation as, for example,
such a property may not be preserved under field redefinitions. While there are no such unambiguous amplitudes at 5- and 6-points for the models we have considered
(other than those fixed to zero by symmetry),
models in which they are present do exist. In particular, the examples studied in
\cite{Wulff:2017vhv} were of this type. If this could be made precise, one
might then hope to either discover a new regularization scheme in which the
integrability is not anomalous or, alternatively, to prove that no such
regularization scheme exists.

%%%%%%%%%%%%%%%%%%%%%%%%%%%%%%%%%%%%%%
\section*{Acknowledgments}
%%%%%%%%%%%%%%%%%%%%%%%%%%%%%%%%%%%%%%

We would like to thank G. Arutyunov, R. Metsaev, R. Roiban, E. Skvortsov and L. Wulff
for useful discussions and comments on the draft.
BH was supported by grant no. 615203 from the European Research Council under the FP7.
NL was supported by the EPSRC grant EP/N509486/1.
AAT was supported by the STFC grant ST/P000762/1.

\bigskip

%%%%%%%%%%%%%%%%%%%%%%%%%%%%%%%%%%%%%%
\appendix
\addcontentsline{toc}{section}{Appendices}
%%%%%%%%%%%%%%%%%%%%%%%%%%%%%%%%%%%%%%

%%%%%%%%%%%%%%%%%%%%%%%%%%%%%%%%%%%%%%
\section[Comments on equivalence of massless S-matrix under field redefinitions]{Comments on equivalence of massless S-matrix under \texorpdfstring{\\}{} field redefinitions}\label{equiv}
%%%%%%%%%%%%%%%%%%%%%%%%%%%%%%%%%%%%%%

Given the on-shell ambiguities discussed in section \ref{kin}, one may wonder
if the massless S-matrix for 2d chiral scalars obeys the standard ``S-matrix
equivalence theorem'' \cite{eqth}. For example, we may consider the general
\sm \rf{2.9} expanded near the trivial vacuum $G_{ab} =\delta_{ab} + {\cal
O}(X)$, $B_{ab}=\O(X)$ and ask if the corresponding S-matrix is invariant under
field redefinitions that preserve the choice of the vacuum and the labelling of
external states. Such redefinitions are of the general form\foot{If the
leading $\l^0 X^a$ term here were also rotated then the scattering states, and
thus the S-matrix, would rotate accordingly. A constant shift of the field
$X^a$ would be a change of the vacuum.}
\be \la{2.311}
X^a \to X^a + \l \, g^a_{\ bc} X^b X^c + \l^2 \, {g}^a_{\ bcd} X^b X^c X^d + \dots \ ,
\ee
where $g^a_{\ bc\dots}$ are constant coefficients.

Let us consider the particular case of two fields $X^a$ ($a=1,2$) and first
compute the tree-level (order $\l^2$) $+- \to +-$ amplitudes. Up to order
$\l^2$, one can write the most general \sm type Lagrangian \rf{2.9} with 20
parameters, and the most general field redefinition \rf{2.31} with 14
parameters. We computed this amplitude and found no dependence on the 14
redefinition parameters. This amplitude has only a Type 1 ambiguity (due to
the diagram in Fig.\ref{figambig} with a vanishing internal momentum when legs
are taken on-shell), which vanishes in both the massive and $i\e$
regularizations.

Next, let us turn to the 6-point amplitudes which are of order $\l^4$ and which
may, in general, contain both Type 1 and Type 2 ambiguities. If we restrict
consideration to the subclass of \sms with only even interactions (which has 32
parameters up to order $\l^4$) then there is a 20-parameter family of field
redefinitions that preserve this property. For these models, the
$\{+,+,+,-,-,-\}$ 6-point amplitudes feature only Type 1 ambiguities and we
again found them to be invariant under field redefinitions in both the massive
and $i\e$-regularizations. Thus both of these regularizations are consistent
with the equivalence theorem in the presence of Type 1 ambiguities.

However, for other amplitudes such as $\{+,+,+,+,-,-\}$, and for generic
6-point amplitudes in models with odd powers of $X$ in the interaction terms,
there are Type 2 ambiguities and they happen to change under field
redefinitions if defined using either the massive or $i \e$-regularization.

This suggests at least two alternatives: (i) these regularizations are not
sufficient for diagrams with Type 2 ambiguities and these require some
additional treatment to satisfy the equivalence theorem; (ii) the standard
S-matrix equivalence theorem may not actually apply to massless amplitudes
involving 2d chiral scalar states. One reason for the latter possibility may
be that the field redefinition \rf{2.31} involves the full field rather than
its chiral parts $X_+$ and $X_-$. Thus perhaps, rather than the chiral
amplitudes themselves being invariant under field redefinitions, only some
special combinations of them may be invariant, describing scattering of the
full field $X$.

One can see that the naive $i \e$-regularization is in tension with the
equivalence theorem as follows. A field redefinition in the free part of the
action produces a vertex involving $\Box X^a$, and when contracted with a
propagator $\Box^{-1}$, this leads to $\Box\times \Box^{-1} $ terms in the
amplitudes (or $k^2/k^2$ in momentum space). For the equivalence theorem to
work these should be resolved as a delta-function (or 1 in momentum space),
while the use of the $i \e$-regularization in the context of massless scalar
scattering (when $k^2$ may go to zero) would typically set such terms to zero.

On the other hand, the massive regularization, where we explicitly deform the
Lagrangian with the same mass term for all fields, i.e. $\L(X) \to \L_m (X)
\equiv \L(X) - \ha m^2 X^2, \ X^2 = X^a X^a$, is naturally a stronger candidate
for satisfying the equivalence. Since massive 2d theories do not feature the
``0/0'' ambiguities, the standard equivalence theorem certainly holds for
$m^2>0$. However, issues arise in taking the massless limit. Suppose two
massless theories $\L$ and $\L'$ are related by a field redefinition $X \to
X'(X)$, i.e. $\L (X) = \L' (X')$. Then their mass-regularized counterparts are
related by $\L'_m (X') = \L_m(X) - \ha m^2 (X'^2 - X^2) $. Thus it is not
$\L_m$ and $\L'_m$ that are related by this field redefinition, but rather
$\L'_m$ is related to $\L_m$ plus the additional vertices, $- \ha m^2 (X'^2 -
X^2)$, which come from the mass term $m^2 X^2$ after the field redefinition.
In the massless limit, these vertices formally vanish being proportional to
$m^2$. However, they may still lead to non-zero contributions to the amplitudes
when multiplied by divergent propagators as internal lines go on-shell in the
massless limit. Thus it is not guaranteed that the massless limit of the
amplitudes computed from $\L_m$ and $\L'_m$ will coincide.

However, it is plausible that in cases with global symmetry, special field
redefinitions that respect this symmetry may be invariances of the
mass-regularized S-matrix (e.g. extra contributions from mass terms may be
forbidden for symmetry reasons).

%v3
Indeed, we have explicitly confirmed that all such globally symmetric redefinitions are non-anomalous for the 6-point amplitudes computed in section \ref{6point}. Starting from the $SO(N+1)/SO(N)$ coset sigma model \rf{2445}, let us consider the most general $SO(N)$-symmetric field redefinition
\be\la{a44}
X^a \to \big(1 + \a \l^2 X^2 + \b \l^4 X^4 + \O(\l^6)
\big)\, X^a \ ,
\ee
specified, up to order $\l^4$, by the two parameters $\a,\b$. Using the massive regularization we find that the
scattering amplitudes in
the resulting theory $\L _{\a, \b}$ are unchanged from their $\a=\b=0$ values \rf{3.10}, \rf{244}.\foot{This is not the case in the $i\e$-regularization.} This follows due to
non-trivial cancellations between the contact diagrams
\begin{align}
A_{\rm cont}^{(+- \to ---+)} &= -\tfrac{i}{16} \l^4 r_+ (k_- + l_- + 2 v_-) (1 + 32 \a + 64 \a^2) \d_{ab} \d_{cd} \d_{ef} \\
&\qquad \qquad\qquad \quad+(\text{other tensor structures}) \ , \no \\
A_{\rm cont}^{(+- \to --++)} &= -\tfrac{i}{16} \l^4 (r_+ + v_+) (k_- + l_-)(1 + 32 \a + 64 \a^2) \d_{ab} \d_{cd} \d_{ef} \\
&\qquad \qquad\qquad \quad+(\text{other tensor structures}) \ , \no
\end{align}
and exchange diagrams.
\foot{In more detail, using that the equivalence theorem certainly applies to the massive theory (before $m$ is sent to zero), one needs only to compute the contribution of new $m^2$-vertices that appear from the $m^2 X^2$ term
upon the field redefinition \rf{a44}.}
Via such symmetric redefinitions, one can reach, e.g., the $SU(2)$ PCM \rf{2.1}-\rf{2.3} in the $N=3$ case ($\a = -{1\ov24}, \, \b = {1\ov 1920}$), and also the alternative coordinates \rf{eq:onsm} for the coset space considered in Appendix \ref{comments} ($\a = -{1\ov 16},\, \b = {1\ov 256}$).

To conclude, there may be anomalies in the equivalence theorem in the case of
massless 2d S-matrices of chiral scalars; we note that somewhat similar issues
appear also in the 4d case when one considers scattering of chiral gauge
vectors \cite{eqvec}. However the massive regularization may exhibit equivalence under special symmetry preserving field redefinitions.

%%%%%%%%%%%%%%%%%%%%%%%%%%%%%%%%%%%%%%
\section{Doubled action for 2d \texorpdfstring{\sms}{sigma-models }and duality symmetry}\label{doubled}
%%%%%%%%%%%%%%%%%%%%%%%%%%%%%%%%%%%%%%

Below we shall review the doubled approach used in section \ref{Tduality},
which was previously applied to computation of massless scalar scattering
amplitudes in \cite{rt}.

A free massless scalar is equivalent to the sum of left and right scalars that
appear as asymptotic states. As for self-dual forms in higher dimensions, the
left and right scalars are independent representations of the 2d Lorentz group
and it is natural to start with an action where each of them is described by an
independent off-shell field. Starting with a Lagrangian $\L(x) \equiv
\L(x,x',\dot{x})$ (where $x'=\del_1 x,\ \dot x = \del_0 x$) we may first put it
into an equivalent phase space form $\hat \L(x,x',p) = p \dot x - H(x, x', p) $
with $p_n = \frac{\del \L}{\del \dot x^n}$. One can then replace the momentum
$p_n$ by another field as $p_n = \del_1 \tx_n$, ending up with the doubled
Lagrangian $\hat \L(x, \tx)$ \cite{t1}. Integrating out $\tx$ gives back the
original path integral for $\L(x)$. We may then replace $x,\tx$ by $x_\pm =
\frac{1}{2}(x \pm \tx)$, which represent the chiral scalars in the free-theory
approximation.

Our focus will be on generic bosonic 2d $\s$-model with
\be \la{1} \L=-\ha \Big( G_{mn} \del^\m x^m \del_\m x^n + \ep^{\m\n} B_{mn} \del_\m x^m \del_\n x^n\Big) = \ha G_{mn} ( \dot x^m \dot x^n - x'^m x'^n) - B_{mn} \dot x^m x'^n\ .
\ee
We shall use the notation $\s^\m=(\s^0, \s^1) \equiv (\tau, \s)$, $\del_0 x =
\dx, \ \del_1 x = x'$, $(\del_\m x)^2 =- \dot x^2 + x'^2$\ \ ($\m,\n=0,1$;\
$m,n=1,\dots,d$). Writing the action for \rf{1} in the ``doubled'' form we get
\cite{t1}
\be
&& \hat S(x,\tx)
= - \ha \int d ^2\s \Big(
- \Omega_{IJ} \dot X^I X'^J + M_{IJ} X'^I X'^J \Big) \ ,
\la{2} \\
&& X=\begin{pmatrix} x \cr \tilde x \end{pmatrix} \ , \ \ \ \ \ \
\Omega = \begin{pmatrix} 0 & I\cr I & 0 \end{pmatrix}\ , \ \ \ \ \ \ \
M = \begin{pmatrix} G- B G^{-1} B & B G^{-1} \cr - G^{-1} B & G^{-1} \end{pmatrix}\ , \la{3}
\ee
where $I,J=1,\dots,2d$ and we have used integration by parts. Explicitly, the
doubled counterpart of the Lagrangian \rf{1} in \rf{2} is
\be \la{4} \hat \L
= \ha ( \dot x^n \tx'_n + \dot \tx_n x'^n) -
\ha (G_{mn} - B_{mk} G^{kl} B_{ln} )
x'^m x'^n - \ha G^{mn} \tx'_m \tx'_n + B_{mk} G^{kn} x'^m \tx'_n\ .
\ee
Starting directly with \rf{2}, $M_{IJ} $ could be a function of the doubled
coordinates $X^I$ \cite{t1} but, in the special case when $G$ and $B$ depend
only on $x^m$, integrating $\tx_m$ out gives back the original, manifestly
Lorentz invariant \sm \rf{1}. Indeed, the doubled theory \rf{4} with $G$ and
$B$ depending only on $x^m$ has Lorentz invariance on shell \cite{t1}. If the
original \sm is integrable (i.e. admits a Lax representation) then the same
will be true also for its doubled counterpart (cf. examples in
\cite{rtw,brtw,rt}).

In the case of $d$ isometric coordinates $x^m$ with the couplings $G,B$
depending only on spectator coordinates and not on $X^I=(x^m, \tx^m)$, the
action \rf{2} is manifestly invariant under the $O(d,d)$ duality
transformations \cite{t1}
\be \la{du}
X'= \La X, \qquad M'= \La^T M \La, \qquad \La^T \Omega \La= \Omega, \qquad \ \ \La\in O(d,d) \ . \ee
Let us also note that the gauge transformations of the $B$-field ($B \to B + d
\xi$, \ $\xi=\xi(x)$), under which the original Lagrangian \rf{1} changes by a
total derivative, remain a symmetry of the doubled action \rf{2} provided one
transforms at the same time the dual coordinate $\tx^m$ (with $x^m$ unchanged).
Indeed, according to the equation of motion $p_m \equiv \tx'_m= G_{mn} \dx^n -
B_{mn} x'^n $, $\tx$ should transform as $\del_1 \tx_m \to \del_1 \tx_m -
(\del_m \xi_n -\del_n \xi_m) \del_1 x^n $, i.e. $\tx_m \to \tx_m + \xi_m -
({\del_1})^{-1} (\del_m \xi_n \del_1 x^n) $.

We shall assume that $G$ has a perturbative expansion near the flat metric,
i.e. $G_{mn} = \delta_{mn} + \O(x)$. Using $ \delta_{mn}$ to raise/lower
indices, let us introduce the combinations $x_\pm$
\be x^m = \xp^m + \xm^m \ , \ \ \ \ \ \ \tx^m = \xp^m -\xm^m \ , \ \ \ \ \ \ \ \ \ \ \ \ \ \ \
x_\pm^m = \tfrac{1}{2} (x^m \pm \tx^m)
\la{5}\ . \ee
Then the free part of the doubled action becomes ($\del_\pm = \pm \del_0 + \del_1$)
\be
\hat \L_0(\xp,\xm)= - \del_1 \xp^n \del_- \xp^n - \del_1 \xm^n \del_+ \xm^n \ .
\la{6} \ee
Thus $x_\pm$ represent chiral scalars on-shell \cite{fj}: the free classical equations are equivalent to
\be \la{7}
\del_- \xp^n=0 \ , \qquad \qquad \del_+ \xm^n=0 \ ,\ee
provided we assume the boundary conditions
\be \la{8}
\del_\mp x_\pm^n\Big|_{|\s|\to \infty} =0 \ee
such that \rf{7} is satisfied at spatial infinity. These are the natural
conditions for discussing the scattering of chiral scalars and they also ensure
the on-shell Lorentz symmetry.\foot{The free action corresponding to \rf{6} is
invariant under the Lorentz-type symmetry: $ \delta x = \tau x' + \sigma \tx',
\ \ \delta \tx = \tau \tx'+ \sigma x',$ or $\delta x_\pm= (\tau \pm \s)
x'_\pm$. An analog of this symmetry exists also in the full interacting action.
This symmetry becomes the standard Lorentz symmetry on the equations of motion
(see \cite{t1} for details).} The on-shell S-matrix elements constructed using
the action for the independent $\xp,\xm$ fields in \rf{4} will then also be
Lorentz invariant. This was explicitly checked on examples in \cite{rt} (cf.
also section \ref{Tduality}).

An advantage of the doubled formalism is that it implies an equivalence
between S-matrices of duality-related models.
The simplest example is provided by the standard T-duality.
Let $x$ be an isometry and $z$ an extra spectator
field. Then the doubled Lagrangian for $x$
is a special case of \rf{4}:
\be \la{44} \hat \L
= \ha ( \dot x \tx' + \dot \tx x') -
\ha G(z)\, x' x' - \ha G^{-1} (z)\, \tx' \tx' -\ha \del^a z \del_a z \ .
\ee
The original theory for the metric $ds^2 = dz^2 + G(z) dx^2$ and its T-dual
$d\td s^2 = dz^2 + G^{-1} (z) d\tx^2$ are related by the inversion of the
coupling $G \to G^{-1}$.\foot{Assuming $G(z) = 1+ c_1 z + c_2 z^2 + \dots$, the
two theories are related by $c_1 \to - c_1, \ c_2 \to - c_2 + c_1^2,$ etc.}
Then the doubled actions for the two dual theories are related simply by $x\to
\tx, \ \tx \to x $, or equivalently\foot{This is just a special case of the
$O(d,d)$ transformation \rf{du} with $\La=\Omega$, or, in the rotated $x_\pm$
basis, $\La= {\rm diag} (I, -I)$. In a $d$-isometric case the doubled action
\rf{2} with couplings $G,B$ not depending on $x^n$ is invariant under the
$O(d,d)$ transformations \rf{du}. Thus with the definition of $x_\pm$ as in
\rf{5} the corresponding S-matrices should remain in direct correspondence but
will be related in a less trivial way than \rf{55} -- via an $O(d,d)$ map
between particular amplitudes.}
\be \la{55}
x_+ \to x_+ \ , \ \ \ \ \qquad x_- \to - x_- \ . \ee
The perturbative S-matrices of the T-dual theories are thus related by \rf{55},
which just amounts to a change of sign of S-matrix elements with an odd number
of $x_-$ fields.\foot{In some special cases (like $G(z)= e^{ \g z}$) the
transformation $G\to G^{-1}$ is equivalent to a simple coordinate redefinition
(like $z\to -z$). In such a case the scattering amplitudes computed in the
doubled approach should be manifestly symmetric under $x_- \to - x_-,\ z \to -
z$ (see \cite{rt}).}

Let us mention that a construction of a doubled action that: (i) reduces to the
original one upon Gaussian integration over half of the fields, and (ii)
describes independent left and right scalars at the free level, may not be
unique. For example, starting with the PCM Lagrangian \rf{2.1}\foot{Here we set
$\l=1$ and use $x^a$ instead of $X^a$ as coordinates on the group.}
\be \la{81}
\L= - \tr \big(J_0^2 - J_1^2\big)\ , \ \ \ \qquad J_\m = g^{-1} \del_\m g \ , \ee
we may construct a doubled model using explicit coordinates as above, i.e. set
$g= e^{t_a x^a}$ and define the dual field via $\del_1 \tx_a= p_a = \frac{\del
\L}{\del_0 x^a} $. But we may also use an alternative definition based on the
algebra-valued momentum conjugate to $ \del_0 g g^{-1} $, i.e. define $\tx $
via $\del_1 \tx = p = {\frac{\del \L}{\del (\del_0 g g^{-1})} } $. In the latter
case the doubled Lagrangian will be
\be \la{9}
\hat \L= - \tr \Big[ 2 \del_1 \tx\, \del_0 g g^{-1} - (\del_1 \tx)^2 - (g^{-1} \del_1 g ) ^2 \Big]
\ . \ee
Similar questions can be asked in the case of the non-abelian dual of PCM.
Starting with the Lagrangian of the NAD model \rf{2.15} originating from the
first-order Lagrangian \rf{2.14}, one can construct the corresponding doubled
action using the general procedure discussed above (see \rf{4}). The momentum
corresponding to the Lagrange multiplier field $Y$ is $J_1$ as it multiplies
$\del_0 Y$ in \rf{2.14}. On the other hand, if we only integrate out $J_0$
from the first-order action \rf{2.14}, then the resulting Lagrangian will
depend only on $Y$ and its momentum $J_1$. We may then replace this momentum
by the new group-valued ``doubled'' variable $g$ as $J_1= g^{-1} \del_1 g $,
ending up with
\be \la{11}
\hat \L =- \tr \Big[ 2 \del_0 Y (g^{-1} \del_1 g) - (g^{-1} \del_1 g)^2 - \big(\del_1 Y - [Y,g^{-1} \del_1 g]\big)^2\Big] \ .
\ee
Identifying $Y\equiv \tx$, we observe that \rf{11} is related to the
alternative doubled Lagrangian for PCM in \rf{9} by the field redefinition $\tx
\to g^{-1} \tx g$.

%%%%%%%%%%%%%%%%%%%%%%%%%%%%%%%%%%%%%%
\section{Massless scattering in \texorpdfstring{$SU(n)$}{SU(n)} PCM: comments on ref. \texorpdfstring{\cite{Gabai:2018tmm}}{[14]}}\label{comments}
%%%%%%%%%%%%%%%%%%%%%%%%%%%%%%%%%%%%%%

The aim of this Appendix is to explain how the discussion of massless PCM
scattering in section 3.2 of \cite{Gabai:2018tmm} is consistent with the
non-vanishing particle production amplitudes found in section \ref{6point} (see
also the discussion in the Introduction). Our starting point is the $SU(n)$
PCM expanded to sextic order in fields as in \cite{Gabai:2018tmm}\foot{Note
that in \cite{Gabai:2018tmm} the authors considered the $U(n)$ PCM but for our
purposes the decoupled $U(1)$ is not relevant. The coupling $\lambda$ is
related to the coupling $g_4$ in \cite{Gabai:2018tmm} as $g_4 =
\frac{\lambda^2}{16} .$ }
\begin{equation}\label{eq:sunpcm}
\mathcal{L} = \tfrac12 \tr[\partial_\mu h \partial^\mu h] + \tfrac{\lambda^2}{8} \tr[h^2 \partial_\mu h \partial^\mu h] + \tfrac{\lambda^4}{64} \tr[h^4 \partial_\mu h \partial^\mu h] + \tfrac{\lambda^4}{128} \tr[h^2 \partial_\mu h h^2 \partial^\mu h ] + \dots ,
\end{equation}
with $h \in \text{Lie}(SU(n))$.
This corresponds to taking $g$ in eq.\eqref{2.1} to be
\begin{equation}
g = \frac{1+\frac{\lambda}{2\sqrt{2}} h}{1-\frac{\lambda}{2\sqrt{2}} h}\ .
\end{equation}
We will be interested in the case of $N=2$, that is the $SU(2)$ PCM or, equivalently,
the $SO(4)/SO(3)$ coset sigma model.
Indeed, setting\foot{Note that the generators $T_a$ used in this Appendix are normalized differently compared to $t_a$ in section 2:
$\tr (T_a T_b) = - \delta_{ab}$.}
\begin{equation}
h = x^a T_a , \qquad T_a = - \tfrac{i}{\sqrt{2}} \sigma_a , \qquad x^2 = x^a x^a , \qquad a = 1,2,3 ,
\end{equation}
where $\sigma_a$ are the Pauli matrices, we find
\begin{equation}\label{eq:onsm}
\mathcal{L} = - \tfrac12 \partial_\mu x^a \partial^\mu x^a + \tfrac{\lambda^2}{16} x^2 \partial_\mu x^a \partial^\mu x^a - \tfrac{3 \lambda^4}{512} (x^2)^2 \partial_\mu x^a \partial^\mu x^a + \dots ,
\end{equation}
which agrees with the expansion of the $SO(4)/SO(3)$ model in \eqref{2445} with
$X^a = \frac{x^a}{1+\frac{\lambda^2}{16} x^2} .$

In this Appendix we are using different target space coordinates compared to those in sections \ref{tree} and \ref{higher}.
Given the subtleties with the equivalence theorem discussed in Appendix \ref{equiv}, it is not a priori clear that amplitudes computed using
the coordinates \eqref{eq:onsm} will agree with those found in sections \ref{tree} and \ref{higher}.
Our focus will be on the particular amplitude \eqref{3.10} for the $SU(2)$ PCM (or equivalently \eqref{246} for the $SO(4)/SO(3)$ model).
As the three Lagrangians in \eqref{2.2}, \eqref{2.3}, in \eqref{2445} and in \eqref{eq:onsm} are
related by $SO(3)$-symmetric field redefinitions,
we expect this amplitude to agree in the three cases when we use the massive regularization
(though it may be different for the $i \eps$-regularization).

In section 3.2 of \cite{Gabai:2018tmm} the authors consider a theory of Hermitian matrix-valued massless fields with 2-derivative interactions.
The claim of \cite{Gabai:2018tmm} is that imposing no tree-level particle production leads to the action of the $U(n)$ PCM.
However, in order to achieve this, the definition of no particle production is weakened.

An $m$-point amplitude in the $U(n)$ or $SU(n)$ PCM can be decomposed in terms of partial colour-ordered amplitudes weighted by the appropriate colour factors
\begin{align}\label{eq:pcoa}
&\mathcal{A}_{a_1 a_2 \dots a_m}(k^{(1)},k^{(2)},\dots,k^{(m)}) \\
&\qquad \qquad \qquad \qquad = \sum_{\sigma \in S_m / \mathbb{Z}_m} \tr[T_{a_{\sigma(1)}} T_{a_{\sigma(2)}} \dots T_{a_{\sigma(m)}} ]\, \mathcal{A}(k^{(\sigma(1))}, k^{(\sigma(2))} , \dots , k^{(\sigma(m))} )\ , \no
\end{align}
where the quotient by $\mathbb{Z}_m$ ensures that we do not double-count cyclic
permutations. Requiring that amplitudes exhibiting particle production should
vanish for all $n$ implies that all the corresponding partial colour-ordered
amplitudes should be zero. In \cite{Gabai:2018tmm} this condition is weakened
to requiring that only partial colour-ordered amplitudes appearing in the sum
in \rf{eq:pcoa} which are \emph{not} of the form
\begin{equation}\label{eq:tpform2}
\mathcal{A}(k^{(1)}_+ , \dots , k^{(m_+)}_+ , l^{(1)}_- , \dots , l^{(m_-)}_- ) \ , \qquad m_+ + m_- = m \ ,
\end{equation}
or cyclic permutations thereof, should vanish. Note that in \eqref{eq:tpform2}
the momenta can be either incoming or outgoing. Physical justifications for
this prescription are given in \cite{Gabai:2018tmm}, however, it is worth
noting that it is unclear how to generalise it to theories that do not have
partial colour-ordered amplitudes, for example, the $SO(N+1)/SO(N)$ coset sigma
model for general $N$.

As the full tree-level amplitude is given by the sum over the various
colour-ordered contributions \eqref{eq:pcoa}, including those of the form
\eqref{eq:tpform2}, it is already clear that the full amplitude may exhibit
particle production even when the prescription of \cite{Gabai:2018tmm} is
satisfied.

Let us consider the particular example
\begin{equation}\label{eq:parex}
\mathcal{A}_{112222}( r_+ , q_- , - k_- , -l_- , - v_- , -r_+ ) \ , \qquad q_- = k_- + l_- + v_- \ ,
\end{equation}
for the $SU(2)$ principal chiral model, which was also computed in section \ref{6point} using different coordinates.
This is the sum of 120 partial colour-ordered amplitudes weighted by the appropriate colour factors, as in eq.\eqref{eq:pcoa}.
To proceed, we need the colour-ordered Feynman rules corresponding to \eqref{eq:sunpcm} (see Fig.\ref{figcolord}).
\begin{figure}
\centering
\begin{tikzpicture}[baseline,scale=0.75,decoration={markings,mark=at position 0.57 with {\arrow{triangle 60}}}]
\draw[-,thick,postaction={decorate}] (-1.4,0.1)--(1.4,0.1);
\node at (0,0.6) {$k$};
\end{tikzpicture}
\qquad
\begin{tikzpicture}[baseline,scale=0.75,decoration={markings,mark=at position 0.57 with {\arrow{triangle 60}}}]
\draw[-,thick,postaction={decorate}] (1,1) -- (0,0);
\draw[-,thick,postaction={decorate}] (-1,1) -- (0,0);
\draw[-,thick,postaction={decorate}] (-1,-1) -- (0,0);
\draw[-,thick,postaction={decorate}] (1,-1) -- (0,0);
\node[left,below] at (-1,-1) {$k^{(1)}$};
\node[right,below] at (1,-1) {$k^{(2)}$};
\node[right,above] at (1,1) {$k^{(3)}$};
\node[left,above] at (-1,1) {$k^{(4)}$};
\end{tikzpicture}
\qquad
\begin{tikzpicture}[baseline,scale=0.75,decoration={markings,mark=at position 0.57 with {\arrow{triangle 60}}}]
\draw[-,thick,postaction={decorate}] (1.4,0) -- (0,0);
\draw[-,thick,postaction={decorate}] (0.7,1.2) -- (0,0);
\draw[-,thick,postaction={decorate}] (-0.7,1.2) -- (0,0);
\draw[-,thick,postaction={decorate}] (-1.4,0) -- (0,0);
\draw[-,thick,postaction={decorate}] (-0.7,-1.2) -- (0,0);
\draw[-,thick,postaction={decorate}] (0.7,-1.2) -- (0,0);
\node[left,below] at (-0.7,-1.2) {$k^{(1)}$};
\node[right,below] at (0.7,-1.2) {$k^{(2)}$};
\node[right] at (1.4,0) {$k^{(3)}$};
\node[right,above] at (0.7,1.2) {$k^{(4)}$};
\node[left,above] at (-0.7,1.2) {$k^{(5)}$};
\node[left] at (-1.4,0) {$k^{(6)}$};
\end{tikzpicture}
\caption{\small The colour-ordered Feynman rules.}
\label{figcolord}
\end{figure}
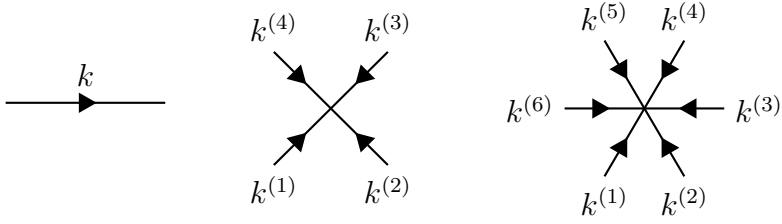
\begin{equation}\begin{split}\la{c8}
P & =\frac{i}{k^2} \ , \qquad V^{(4)} = \tfrac{i}{8} \lambda^2 (k^{(1)} + k^{(3)})^2 = \tfrac{i}{8}\lambda^2 (k^{(2)} + k^{(4)})^2 \ ,
\\
V^{(6)} & = \tfrac{i}{64} \lambda^4 (k^{(1)} + k^{(3)} + k^{(5)})^2 = \tfrac{i}{64} \lambda^4 (k^{(2)} + k^{(4)} + k^{(6)})^2 \ .
\end{split}\end{equation}
The partial colour-ordered six-point amplitude is then given by the sum of the four graphs in Fig.\ref{colordgraphs}.
\begin{figure}
\centering
\begin{tikzpicture}[baseline,scale=0.75,decoration={markings,mark=at position 0.57 with {\arrow{triangle 60}}}]
\draw[-,thick,postaction={decorate}] (1.4,0) -- (0,0);
\draw[-,thick,postaction={decorate}] (0.7,1.2) -- (0,0);
\draw[-,thick,postaction={decorate}] (-0.7,1.2) -- (0,0);
\draw[-,thick,postaction={decorate}] (-1.4,0) -- (0,0);
\draw[-,thick,postaction={decorate}] (-0.7,-1.2) -- (0,0);
\draw[-,thick,postaction={decorate}] (0.7,-1.2) -- (0,0);
\node[left,below] at (-0.7,-1.2) {$k^{(1)}$};
\node[right,below] at (0.7,-1.2) {$k^{(2)}$};
\node[right] at (1.4,0) {$k^{(3)}$};
\node[right,above] at (0.7,1.2) {$k^{(4)}$};
\node[left,above] at (-0.7,1.2) {$k^{(5)}$};
\node[left] at (-1.4,0) {$k^{(6)}$};
\end{tikzpicture} \qquad
\begin{tikzpicture}[baseline,scale=0.5,decoration={markings,mark=at position 0.57 with {\arrow{triangle 60}}}]
\draw[-,thick,postaction={decorate}] (-1.4,0.7) -- (0,0.7);
\draw[-,thick,postaction={decorate}] (1.4,0.7) -- (0,0.7);
\draw[-,thick,postaction={decorate}] (-1.4,-0.7) -- (0,-0.7);
\draw[-,thick,postaction={decorate}] (1.4,-0.7) -- (0,-0.7);
\draw[-,thick,postaction={decorate}] (0,2.1) -- (0,0.7);
\draw[-,thick,postaction={decorate}] (0,-2.1) -- (0,-0.7);
\draw[-,thick] (0,0.7) -- (0,-0.7);
\node[left] at (-1.4,-0.7) {$k^{(1)}$};
\node[below] at (0,-2.1) {$k^{(2)}$};
\node[right] at (1.4,-0.7) {$k^{(3)}$};
\node[right] at (1.4,0.7) {$k^{(4)}$};
\node[above] at (0,2.1) {$k^{(5)}$};
\node[left] at (-1.4,0.7) {$k^{(6)}$};
\end{tikzpicture} \qquad
\begin{tikzpicture}[baseline,scale=0.5,decoration={markings,mark=at position 0.57 with {\arrow{triangle 60}}}]
\draw[-,thick,postaction={decorate}] (-1.4,0.7) -- (0,0.7);
\draw[-,thick,postaction={decorate}] (1.4,0.7) -- (0,0.7);
\draw[-,thick,postaction={decorate}] (-1.4,-0.7) -- (0,-0.7);
\draw[-,thick,postaction={decorate}] (1.4,-0.7) -- (0,-0.7);
\draw[-,thick,postaction={decorate}] (0,2.1) -- (0,0.7);
\draw[-,thick,postaction={decorate}] (0,-2.1) -- (0,-0.7);
\draw[-,thick] (0,0.7) -- (0,-0.7);
\node[left] at (-1.4,-0.7) {$k^{(2)}$};
\node[below] at (0,-2.1) {$k^{(3)}$};
\node[right] at (1.4,-0.7) {$k^{(4)}$};
\node[right] at (1.4,0.7) {$k^{(5)}$};
\node[above] at (0,2.1) {$k^{(6)}$};
\node[left] at (-1.4,0.7) {$k^{(1)}$};
\end{tikzpicture} \qquad
\begin{tikzpicture}[baseline,scale=0.5,decoration={markings,mark=at position 0.57 with {\arrow{triangle 60}}}]
\draw[-,thick,postaction={decorate}] (-1.4,0.7) -- (0,0.7);
\draw[-,thick,postaction={decorate}] (1.4,0.7) -- (0,0.7);
\draw[-,thick,postaction={decorate}] (-1.4,-0.7) -- (0,-0.7);
\draw[-,thick,postaction={decorate}] (1.4,-0.7) -- (0,-0.7);
\draw[-,thick,postaction={decorate}] (0,2.1) -- (0,0.7);
\draw[-,thick,postaction={decorate}] (0,-2.1) -- (0,-0.7);
\draw[-,thick] (0,0.7) -- (0,-0.7);
\node[left] at (-1.4,-0.7) {$k^{(3)}$};
\node[below] at (0,-2.1) {$k^{(4)}$};
\node[right] at (1.4,-0.7) {$k^{(5)}$};
\node[right] at (1.4,0.7) {$k^{(6)}$};
\node[above] at (0,2.1) {$k^{(1)}$};
\node[left] at (-1.4,0.7) {$k^{(2)}$};
\end{tikzpicture}
\caption{\small Contributions to the partial colour-ordered six-point amplitude.}\label{colordgraphs}
\end{figure}
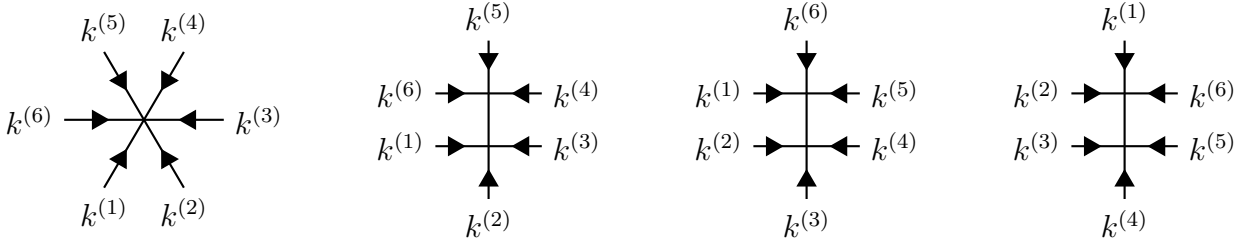
%%%%%%%%%%%%%%%%%%%%%%%%%%%%%%%%%%%%%%
\begin{table}[t!]
\begin{tabular}{|c|c|c|}
\hline
& Colour factor & Colour-ordered amplitude
\\\hline\hline
$\mathcal{A}(r_+,-r_+,q_-,-k_-,-l_-,-v_-)+ \mathrm{perms}$
&
$\tr[T_1 T_2 T_1 T_2^3] = \frac14$
&
$ \tfrac{(3-4\omega) i}{64} \lambda^4 r_+ q_-$
\\\hline
$\mathcal{A}(r_+,-r_+,-k_-,q_-,-l_-,-v_-)+ \mathrm{perms}$
&
$\tr[T_1 T_2^2 T_1 T_2^2] = - \frac14$
&
$-\tfrac{(1-4\omega)i}{64} \lambda^4 r_+ q_-$
\\\hline
$\mathcal{A}(r_+,-r_+,-k_-,-l_-,q_-,-v_-)+ \mathrm{perms}$
&
$\tr[T_1 T_2 T_1 T_2^3] = \frac14$
&
$\tfrac{(1-4\omega)i}{64} \lambda^4 r_+ q_-$
\\\hline
$\mathcal{A}(r_+,-r_+,-k_-,-l_-,-v_-,q_-)+ \mathrm{perms}$
&
$\tr[T_1^2 T_2^4] = - \frac14$
&
$- \tfrac{(3-4\omega) i}{64} \lambda^4 r_+ q_-$
\\\hline\hline
$\mathcal{A}(r_+,q_-,-r_+,-k_-,-l_-,-v_-)+ \mathrm{perms}$
&
$\tr[T_1^2 T_2^4] = - \frac14$
&
0
\\\hline
$\mathcal{A}(r_+,-k_-,-r_+,q_-,-l_-,-v_-)+ \mathrm{perms}$
&
$\tr[T_1 T_2^2 T_1 T_2^2] = - \frac14$
&
0
\\\hline
$\mathcal{A}(r_+,-k_-,-r_+,-l_-,q_-,-v_-)+ \mathrm{perms}$
&
$\tr[T_1 T_2 T_1 T_2^3] = \frac14$
&
0
\\\hline
$\mathcal{A}(r_+,-k_-,-r_+,-l_-,-v_-,q_-)+ \mathrm{perms}$
&
$\tr[T_1^2 T_2^4] = - \frac14$
&
0
\\\hline\hline
$\mathcal{A}(r_+,q_-,-k_-,-r_+,-l_-,-v_-)+ \mathrm{perms}$
&
$\tr[T_1^2 T_2^4] = - \frac14$
&
0
\\\hline
$\mathcal{A}(r_+,-k_-,q_-,-r_+,-l_-,-v_-)+ \mathrm{perms}$
&
$\tr[T_1 T_2 T_1 T_2^3] = \frac14$
&
0
\\\hline
$\mathcal{A}(r_+,-k_-,-l_-,-r_+,q_-,-v_-)+ \mathrm{perms}$
&
$\tr[T_1 T_2 T_1 T_2^3] = \frac14$
&
0
\\\hline
$\mathcal{A}(r_+,-k_-,-l_-,-r_+,-v_-,q_-)+ \mathrm{perms}$
&
$\tr[T_1^2 T_2^4] = - \frac14$
&
0
\\\hline\hline
$\mathcal{A}(r_+,q_-,-k_-,-l_-,-r_+,-v_-)+ \mathrm{perms}$
&
$\tr[T_1^2 T_2^4] = - \frac14$
&
0
\\\hline
$\mathcal{A}(r_+,-k_-,q_-,-l_-,-r_+,-v_-)+ \mathrm{perms}$
&
$\tr[T_1 T_2 T_1 T_2^3] = \frac14$
&
0
\\\hline
$\mathcal{A}(r_+,-k_-,-l_-,q_-,-r_+,-v_-)+ \mathrm{perms}$
&
$\tr[T_1 T_2^2 T_1 T_2^2] = - \frac14$
&
0
\\\hline
$\mathcal{A}(r_+,-k_-,-l_-,-v_-,-r_+,q_-)+ \mathrm{perms}$
&
$\tr[T_1^2 T_2^4] = - \frac14$
&
0
\\\hline\hline
$\mathcal{A}(r_+,q_-,-k_-,-l_-,-v_-,-r_+)+ \mathrm{perms}$
&
$\tr[T_1^2 T_2^4] = - \frac14$
&
$- \tfrac{(3-4\omega) i}{64} \lambda^4 r_+ q_-$
\\\hline
$\mathcal{A}(r_+,-k_-,q_--l_-,-v_-,-r_+)+ \mathrm{perms}$
&
$\tr[T_1 T_2 T_1 T_2^3] = \frac14$
&
$\tfrac{(1-4\omega)i}{64} \lambda^4 r_+ q_-$
\\\hline
$\mathcal{A}(r_+,-k_-,-l_-,q_-,-v_-,-r_+)+ \mathrm{perms}$
&
$\tr[T_1 T_2^2 T_1 T_2^2] = - \frac14$
&
$-\tfrac{(1-4\omega)i}{64} \lambda^4 r_+ q_-$
\\\hline
$\mathcal{A}(r_+,-k_-,-l_-,-v_-,q_-,-r_+)+ \mathrm{perms}$
&
$\tr[T_1 T_2 T_1 T_2^3] = \frac14$
&
$\tfrac{(3-4\omega) i}{64} \lambda^4 r_+ q_-$
\\\hline
\end{tabular}
\caption{\small Table of colour-ordered contributions to the amplitude \eqref{eq:parex}.}\label{tabex}
\end{table}%
For the particular tree-level amplitude \eqref{eq:parex} there will be no
ambiguities of Type 1, but there will be Type 2 ambiguities. To implement both
the massive and $i \eps$-regularizations in the same computation, we use the
massive regularization, and weight those graphs with a Type 2 ambiguity with a
factor of $\omega$. This gives the massive regularization result for $\omega =
1$ and the $i \eps$-regularization result for $\omega = 0$.

Now grouping together those partial colour-ordered amplitudes that are related
by permuting $k_-$, $l_-$ and $v_-$, we are left with the 20 terms in Table
\ref{tabex}. In agreement with the prescription of \cite{Gabai:2018tmm}, the
non-vanishing partial colour-ordered amplitudes in rows 1-4 and 17-20 are
precisely those that take the form \eqref{eq:tpform2}. Summing all the
contributions in Table \ref{tabex} we find
\begin{equation}\label{eq:finform}
\mathcal{A}_{112222}(r_+,q_-,-r_+,-k_-,-l_-,-v_-) = \tfrac{i(1-2\omega)}{16} \lambda^4 r_+ q_- \ .
\end{equation}
For the massive regularization, $\omega = 1$, we indeed find agreement with the
corresponding result \eqref{246} from the $SU(2)$ PCM and the $SO(4)/SO(3)$ coset sigma %v2
model.\foot{The form of the amplitude \eqref{eq:finform} might suggest that
choosing $\omega = \frac12$ could provide an alternative to the massive and
$i\eps$-regularizations that leads to no particle production. However, this is
accidental for this particular amplitude and does not work in general. The PCM
exhibits particle production for any $\omega$. Indeed, observe that the
amplitudes \eqref{244} and \eqref{247} also demonstrate particle production,
but as they only involve Type 1 ambiguities they will not depend on $\omega$.
Furthermore, it remains the case that the massive regularization, $\omega = 1$,
is the ``closest'' to being consistent with equivalence under field
redefinitions.}

%v2
It is worth noting that the prescription of \cite{Gabai:2018tmm}
does not appear to be preserved under field
redefinitions. Let us consider a particular class of redefinitions
\begin{equation} \la{sunredef}
h \to h - \tfrac{1}{16} \alpha \lambda^2 h^3 + \tfrac{1}{128} \beta \lambda^4 h^5 \ ,
\end{equation}
in the PCM Lagrangian \eqref{eq:sunpcm}, which then takes the form
\begin{equation}\begin{split}\label{eq:sunpcm2}
\mathcal{L} & = \tfrac12 \tr[\partial_\mu h \partial^\mu h] + \tfrac18(1-\alpha)\lambda^2\tr[h^2\partial_\mu h \partial^\mu h] - \tfrac{1}{16}\alpha\lambda^2\tr[h \partial_\mu h h \partial^\mu h]
\\ & \quad
+ \tfrac1{256}(4-12\alpha+\alpha^2+4\beta)\lambda^4 \tr[h^4\partial_\mu h \partial^\mu h]
+ \tfrac1{128}(-2\alpha+\alpha^2+2\beta)\lambda^4 \tr[h^3 \partial_\mu h h \partial^\mu h]
\\ & \quad + \tfrac1{512}(4+3\alpha^2+4\beta)\lambda^4 \tr[h^2 \partial_\mu h h^2 \partial^\mu h ] + \dots .
\end{split}
\end{equation}
The colour-ordered Feynman rules are now (cf. \rf{c8})
\begin{align}\label{eq:cofr2}
P & =\frac{i}{k^2} \ , \qquad V^{(4)} = \tfrac{i}{8}(1-\alpha)\lambda^2 (k^{(1)}+k^{(3)})^2 + \tfrac{i}{8} \alpha \lambda^2 (k^{(1)}\cdot k^{(3)} + k^{(2)} \cdot k^{(4)})\ ,
\\
V^{(6)} & =
- \tfrac{i}{256} (4+3\alpha^2+4\beta)\lambda^4 (k^{(1)} \cdot k^{(4)} + k^{(2)} \cdot k^{(5)} + k^{(3)} \cdot k^{(6)}) \no
\\
- \tfrac{i}{128} & (-2\alpha+\alpha^2+2\beta)\lambda^4 (k^{(1)} \cdot k^{(3)} + k^{(1)} \cdot k^{(5)} + k^{(3)} \cdot k^{(5)} + k^{(2)} \cdot k^{(4)} + k^{(2)} \cdot k^{(6)} + k^{(4)} \cdot k^{(6)}) \no
\\
- \tfrac{i}{256} &(4-12\alpha+\alpha^2+4\beta)\lambda^4(k^{(1)} \cdot k^{(2)} + k^{(2)} \cdot k^{(3)} + k^{(3)} \cdot k^{(4)} + k^{(4)} \cdot k^{(5)} + k^{(5)} \cdot k^{(6)} + k^{(6)} \cdot k^{(1)} ) \ . \no
\end{align}
\begin{table}[t!]
\begin{tabular}{|c|c|c|}
\hline
& Colour factor & Colour-ordered amplitude
\\\hline\hline
$\mathcal{A}(r_+,-r_+,q_-,-k_-,-l_-,-v_-)+ \mathrm{perms}$
&
$\tr[T_1 T_2 T_1 T_2^3] = \frac14$
&
$\tfrac{i}{64} (3-8\alpha-4\omega(1-\alpha))\lambda^4 r_+ q_-$
\\\hline
$\mathcal{A}(r_+,-r_+,-k_-,q_-,-l_-,-v_-)+ \mathrm{perms}$
&
$\tr[T_1 T_2^2 T_1 T_2^2] = - \frac14$
&
$-\tfrac{i}{64}(1+8\alpha-4\omega) \lambda^4 r_+ q_-$
\\\hline
$\mathcal{A}(r_+,-r_+,-k_-,-l_-,q_-,-v_-)+ \mathrm{perms}$
&
$\tr[T_1 T_2 T_1 T_2^3] = \frac14$
&
$\tfrac{i}{64}(1+8\alpha-4\omega) \lambda^4 r_+ q_-$
\\\hline
$\mathcal{A}(r_+,-r_+,-k_-,-l_-,-v_-,q_-)+ \mathrm{perms}$
&
$\tr[T_1^2 T_2^4] = - \frac14$
&
$-\tfrac{i}{64} (3-8\alpha-4\omega(1-\alpha))\lambda^4 r_+ q_-$
\\\hline\hline
$\mathcal{A}(r_+,q_-,-r_+,-k_-,-l_-,-v_-)+ \mathrm{perms}$
&
$\tr[T_1^2 T_2^4] = - \frac14$
&
0
\\\hline
$\mathcal{A}(r_+,-k_-,-r_+,q_-,-l_-,-v_-)+ \mathrm{perms}$
&
$\tr[T_1 T_2^2 T_1 T_2^2] = - \frac14$
&
$-\tfrac{3i}{16}\alpha\lambda^4 r_+ q_-$
\\\hline
$\mathcal{A}(r_+,-k_-,-r_+,-l_-,q_-,-v_-)+ \mathrm{perms}$
&
$\tr[T_1 T_2 T_1 T_2^3] = \frac14$
&
0
\\\hline
$\mathcal{A}(r_+,-k_-,-r_+,-l_-,-v_-,q_-)+ \mathrm{perms}$
&
$\tr[T_1^2 T_2^4] = - \frac14$
&
$\tfrac{3i}{16}\alpha\lambda^4 r_+ q_-$
\\\hline\hline
$\mathcal{A}(r_+,q_-,-k_-,-r_+,-l_-,-v_-)+ \mathrm{perms}$
&
$\tr[T_1^2 T_2^4] = - \frac14$
&
$\tfrac{i}{16}\alpha\lambda^4 r_+ q_-$
\\\hline
$\mathcal{A}(r_+,-k_-,q_-,-r_+,-l_-,-v_-)+ \mathrm{perms}$
&
$\tr[T_1 T_2 T_1 T_2^3] = \frac14$
&
$-\tfrac{i}{16}\alpha\lambda^4 r_+ q_-$
\\\hline
$\mathcal{A}(r_+,-k_-,-l_-,-r_+,q_-,-v_-)+ \mathrm{perms}$
&
$\tr[T_1 T_2 T_1 T_2^3] = \frac14$
&
$-\tfrac{i}{16}\alpha\lambda^4 r_+ q_-$
\\\hline
$\mathcal{A}(r_+,-k_-,-l_-,-r_+,-v_-,q_-)+ \mathrm{perms}$
&
$\tr[T_1^2 T_2^4] = - \frac14$
&
$\tfrac{i}{16}\alpha\lambda^4 r_+ q_-$
\\\hline\hline
$\mathcal{A}(r_+,q_-,-k_-,-l_-,-r_+,-v_-)+ \mathrm{perms}$
&
$\tr[T_1^2 T_2^4] = - \frac14$
&
$\tfrac{3i}{16}\alpha\lambda^4 r_+ q_-$
\\\hline
$\mathcal{A}(r_+,-k_-,q_-,-l_-,-r_+,-v_-)+ \mathrm{perms}$
&
$\tr[T_1 T_2 T_1 T_2^3] = \frac14$
&
0
\\\hline
$\mathcal{A}(r_+,-k_-,-l_-,q_-,-r_+,-v_-)+ \mathrm{perms}$
&
$\tr[T_1 T_2^2 T_1 T_2^2] = - \frac14$
&
$-\tfrac{3i}{16}\alpha\lambda^4 r_+ q_-$
\\\hline
$\mathcal{A}(r_+,-k_-,-l_-,-v_-,-r_+,q_-)+ \mathrm{perms}$
&
$\tr[T_1^2 T_2^4] = - \frac14$
&
0
\\\hline\hline
$\mathcal{A}(r_+,q_-,-k_-,-l_-,-v_-,-r_+)+ \mathrm{perms}$
&
$\tr[T_1^2 T_2^4] = - \frac14$
&
$-\tfrac{i}{64} (3-8\alpha-4\omega(1-\alpha))\lambda^4 r_+ q_-$
\\\hline
$\mathcal{A}(r_+,-k_-,q_--l_-,-v_-,-r_+)+ \mathrm{perms}$
&
$\tr[T_1 T_2 T_1 T_2^3] = \frac14$
&
$\tfrac{i}{64}(1+8\alpha-4\omega) \lambda^4 r_+ q_-$
\\\hline
$\mathcal{A}(r_+,-k_-,-l_-,q_-,-v_-,-r_+)+ \mathrm{perms}$
&
$\tr[T_1 T_2^2 T_1 T_2^2] = - \frac14$
&
$-\tfrac{i}{64}(1+8\alpha-4\omega) \lambda^4 r_+ q_-$
\\\hline
$\mathcal{A}(r_+,-k_-,-l_-,-v_-,q_-,-r_+)+ \mathrm{perms}$
&
$\tr[T_1 T_2 T_1 T_2^3] = \frac14$
&
$\tfrac{i}{64} (3-8\alpha-4\omega(1-\alpha))\lambda^4 r_+ q_-$
\\\hline
\end{tabular}
\caption{\small Table of colour-ordered contributions to the amplitude \eqref{eq:parex} for the
colour-ordered Feynman rules \eqref{eq:cofr2}.}\label{tabex2}
\end{table}%
Let us again consider the amplitude \eqref{eq:parex}. Grouping together those
partial colour-ordered amplitudes that are related by permuting $k_-$, $l_-$
and $v_-$, we are left with the 20 terms in Table \ref{tabex2}. We see that
for $\alpha \neq 0$ it is no longer the case that the non-vanishing partial
colour-ordered amplitudes all take the form \eqref{eq:tpform2} as assumed in
\cite{Gabai:2018tmm}.
Therefore, while the prescription of \cite{Gabai:2018tmm} does lead to the PCM in
one particular set of coordinates and hence may be used as a constructive procedure,
their definition of no particle production does not seem to be preserved under
field redefinitions.\foot{In Tables \ref{tabex} and \ref{tabex2}, rows 1-8 and
13-20 correspond to amplitudes that contain an IR ambiguity, while rows 9-12 do not
have an ambiguity. Let us note, that the amplitudes in rows 9-12 also change
under field redefinitions even though here there is no IR ambiguity.}
Thus it cannot, in general, be used as a test of integrability.

Finally summing all the contributions in Table \ref{tabex2} we find
\begin{equation}\label{eq:finform2}
\mathcal{A}_{112222}(r_+,q_-,-r_+,-k_-,-l_-,-v_-) = \tfrac{i}{16} \big(1-2\omega - \alpha(1-\omega)\big) \lambda^4 r_+ q_- \ .
\end{equation}
For the massive regularization, $\omega = 1$, the dependence on the field redefinition parameter $\alpha$
drops out as expected (since the field redefinition \eqref{sunredef} is $SU(n)$-symmetric -- see Appendix \ref{equiv}),
and we again obtain agreement with the corresponding
result \eqref{246} from the $SU(2)$ PCM and the $SO(4)/SO(3)$ coset sigma
model.

In conclusion, the discussion in \cite{Gabai:2018tmm} does not actually
contradict the presence of particle production in the PCM that we observed
above. However, let us emphasise that it is unclear how to extend the prescription of ``no
particle production for partial colour-ordered amplitudes'' used in
\cite{Gabai:2018tmm} to general massless integrable theories that do not have a
notion of colour-ordering, for example, to the $SO(N+1)/SO(N)$ coset sigma
model for $N \neq 3$. Furthermore, this prescription does not appear to be preserved
under general field redefinitions.

\bigskip

%%%%%%%%%%%%%%%%%%%%%%%%%%%%%%%%%%%%%%

%%%%%%%%%%%%%%%%%%%%%%%%%%%%%%%%%%%%%%

\end{document}
%%%%%%%%%%%%%%%%%%%%%%%%%%%%%%%%%%%%%%